\documentclass[pra,twocolumn,amsmath,amssymb,unsortedaddress,superscriptaddress,showpacs]{revtex4-1}

\usepackage{latexsym,epsfig,graphicx}
\usepackage{dcolumn}
\usepackage{bm}
\usepackage[colorlinks,urlcolor=blue,citecolor=blue]{hyperref}
\usepackage{comment}

\begin{document}

\title{Static and dynamic properties of shell-shaped condensates}

\author{Kuei Sun}
\thanks{kuei.sun@utdallas.edu}
\affiliation{Department of Physics, The University of Texas at
Dallas, Richardson, Texas 75080-3021, USA}
\author{Karmela Padavi\'{c}}
\affiliation{Department of Physics, University of Illinois at
Urbana-Champaign, Urbana, Illinois 61801-3080, USA}
\author{Frances Yang}
\affiliation{Department of Physics, Smith College, Northampton,
Massachusetts 01063, USA}
\author{Smitha Vishveshwara}
\thanks{smivish@illinois.edu}
\affiliation{Department of Physics, University of Illinois at
Urbana-Champaign, Urbana, Illinois 61801-3080, USA}
\author{Courtney Lannert}
\thanks{clannert@smith.edu}
\affiliation{Department of Physics, Smith College, Northampton,
Massachusetts 01063, USA}
\affiliation{Department of Physics, University of Massachusetts, Amherst, Massachusetts 01003-9300, USA}

\begin{abstract}
Static, dynamic, and topological properties of hollow systems
differ from those that are fully filled as a result of the
presence of a boundary associated with an inner surface. Hollow
Bose-Einstein condensates (BECs) naturally occur in various
ultracold atomic systems and possibly within neutron stars but
have hitherto not been experimentally realized in isolation on
Earth because of gravitational sag. Motivated by the expected
first realization of fully closed BEC shells in the microgravity
conditions of the Cold Atomic Laboratory aboard the International
Space Station, we present a comprehensive study of spherically
symmetric hollow BECs as well as the hollowing transition from a
filled sphere BEC into a thin shell through central density
depletion. We employ complementary analytic and numerical
techniques in order to study equilibrium density profiles and the
collective mode structures of condensate shells hosted by a range
of trapping potentials. We identify concrete and robust signatures
of the evolution from filled to hollow structures and the effects
of the emergence of an inner boundary, inclusive of a dip in
breathing-mode-type collective mode frequencies and a
restructuring of surface mode structure across the transition. By
extending our analysis to a two-dimensional transition of a disk
to a ring, we show that the collective mode signatures are an
essential feature of hollowing, independent of the specific
geometry. Finally, we relate our work to past and ongoing
experimental efforts and consider the influence of gravity on thin
condensate shells. We identify the conditions under which
gravitational sag is highly destructive and study the mode-mixing
effects of microgravity on the collective modes of these shells.
\end{abstract}

\maketitle

\section{Introduction and Motivation}\label{sec:I}

The realization of Bose-Einstein condensation in dilute ultracold
atomic gases gave rise to spectacular directions in testing and
exploring quantum phenomena at macroscopic
scales~\cite{Dalfovo1999,Leggett2001,Cornell2002,Ketterle2002,Pethick2008,Bloch2008}.
Since the advent of this rich field, control and manipulation of
these quantum fluids in diverse trapping potentials have yielded
Bose-Einstein condensates (BECs) in a variety of
geometries~\cite{Gorlitz2001,
Greiner2001,Dettemer2001,Hechenblaikner2005,
Smith2005,Ramanathan2011,Gupta2005, Gaunt2013, Shin2004,Shin2004,
Hofferberth2006, Smerzi1997, Lannert2007}. Here, we consider a
fundamentally different geometry for a BEC---a hollow spherical
shell. We theoretically study its evolution from a filled sphere
to one having a small hollow region at its center to the thin
spherical shell limit. Hitherto, creating hollow spheres has been
an experimental challenge on Earth because of gravitational sag.
Our study is particularly timely as a realization of a
shell-shaped condensate~\cite{Lundblad,Hollow} is expected to take
place within a ``bubble trap"~\cite{Zobay2001} under microgravity
conditions in the space-based Cold Atomic Laboratory~\cite{CAL},
recently launched to provide opportunities to investigate BEC
behaviors that are unobservable in terrestrial
laboratories~\cite{CAL_NatureNews}.

A BEC in a shell-shaped geometry is fundamentally interesting from
multiple perspectives: (i) Tuning between the filled sphere and
thin shell limit offers a means of achieving dimensional
cross-over from three-dimensional (3D) to two-dimensional (2D)
behavior. We expect thermodynamics and collective mode properties
to be acutely sensitive to dimensionality. (ii) The point at which
the sphere initially hollows represents a change in topology. In
contrast to other studies of topological transitions in quantum
fluids, which require measuring Berry phases and related global
invariants in momentum space, this topological change directly
involves physical geometry. The filled and hollow spheres
correspond to different second homotopy groups in that, unlike for
the filled condensate, a spherical surface within the hollow BEC
that surrounds its center cannot be continuously deformed into a
point. (iii) The change in topology has physical consequences as
it is accompanied by the appearance of a new, inner boundary. As
we show here, the boundary has a marked effect on collective-mode
structures. In terms of dimensional crossover and topological
change, the hollowing out of the spherical BEC is a higher
dimensional analog of BEC systems that have recently generated
much interest upon their realization---annular and toroidal
BECs~\cite{Ramanathan2011, Gupta2005}. (iv) The shell BEC also
offers noteworthy features that are not present in these
geometries, such as vortex phenomena when subject to rotation and
associated Kosterlitz-Thouless physics~\cite{Kosterlitz1973} on a
curved, edgeless surface in the thin shell limit.

Shell-shaped BECs appear in a range of strongly correlated system
from the micron to the astronomical scale; the BEC studied here,
being in isolation, provides the simplest such instance and thus
acts as a test bed for more complex situations. As an established
case, shell-shaped condensate regions occur in three-dimensional
(3D) optical lattice systems of ultracold
bosons~\cite{Jaksch1998,Oosten2001,Greiner2002,Greiner2003},
produced by  the interplay between weak tunneling-to-interaction
ratio (or strong lattice
potential)~\cite{Fisher1989,Sheshadri1993,Freericks1994} and
inhomogeneous boson density due to harmonic
confinement~\cite{Batrouni2002,DeMarco2005,Campbell2006}. In the
case of co-existing phases, superfluid (SF) regions are confined
by surrounding Mott-insulator (MI) regions of the same bosons,
creating an effective trapping
potential~\cite{Barankov2007,Sun2009}, rather than by an external
trap. In addition, condensate shells are expected in Bose-Fermi
mixtures~\cite{Molmer1998,Ospelkaus2006,Schaeybroeck2009}, where a
shell of Bose gas results from a phase separation from a core of
fermions. At the astronomical scale, signals from neutron stars
acting as radio pulsars have given rise to models that suggest the
existence of macroscopic quantum states of matter, some possibly
corresponding to shell crusts of neutron
superfluids~\cite{Weber2004,Pethick2015}.

\begin{figure}[t]
\centering
  \includegraphics[width=8.6cm]{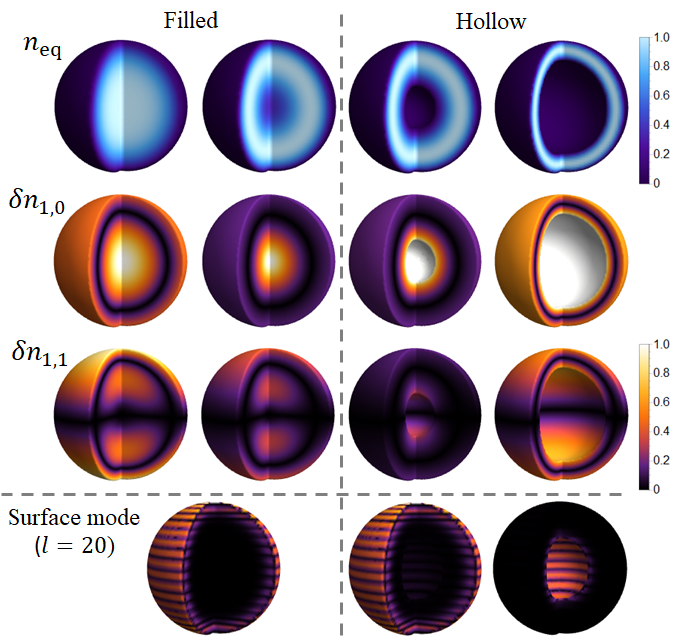}
  \caption{(Color online)
Top row: schematic equilibrium density profiles of a spherical BEC
evolving from filled (left two panels) to hollow (right two)
geometries. The middle two columns are close to the hollowing
transition. Second and third rows: density deviation of collective
modes $(\nu,\ell)=(1,0)$ and $(1,1)$, respectively, for the
corresponding equilibrium density profile in the same column.
Bottom row: density deviation of high angular momentum ($\ell=20$)
surface modes on a filled BEC (left, on the outer surface only)
and a hollow BEC (right, on either the inner or outer surface).
The density and density deviations are scaled by colors in the
corresponding bar graph. }
        \label{fig:1}
\end{figure}

In previous work~\cite{Padavic2016}, we presented preliminary
evidence that a three-dimensional topological hollowing transition
in a BEC should be accompanied by specific signatures in the
collective mode spectra and illustrated these features using
numerical solutions to the hydrodynamic equations for the specific
case of the bubble-trap potential. In this work, we broaden and
generalize the analysis of shell-shaped condensates and the
hollowing transition, and directly address the experimental
feasibility of detecting such a topological change. To that end,
we use a family of trapping potentials (including the bubble trap)
that can continuously tune between the filled sphere and the
thin-shell limits. We perform an analysis of two concrete
condensate properties---density profiles and collective mode
structure, shown schematically in Fig.~\ref{fig:1}, and show that
the two are intertwined during the topological transition in which
the filled condensate becomes hollow. By employing a generalized
trapping potential that allows us to adjust the density profile
near the center of the BEC as it hollows, we are able to
distinguish universal topological hollowing features in the
collective mode spectra from those that are dependent on trap
geometry. Thus, the work presented here provides a deeper
understanding of physical mechanisms that result in signatures of
the topological change in the shape of the BEC. Collective modes
were the first phenomenon to be studied after the successful
production of BECs and are well understood in the filled sphere
case~\cite{Jin1996,Mewes1996,Edwards1996,Stringari1996,Castin1996,Stamper-Kurn1998,Chevy2002,Fort2003,Yang2009,Haller2009,Pollack2010,Kuwamoto2012,Lobser2015,Straatsma2016};
by taking the sphere as one limit in our family of BEC shapes, we
show that the collective mode frequencies in the filled sphere and
thin shell limits reflect the 3D and 2D limiting behaviors of the
hollowing system. We additionally use an \emph{in situ} numerical
simulation of an experimental probe of collective modes to show
that predicted hydrodynamic features in the spectra would in fact
be relevant in experimental settings.

As our main results, we find that the crossover between the filled
sphere and thin shell limits has several rich features and that
the collective mode spectra in the system are excellent probes of
the transition from filled to hollow geometry, in that: (i) the
mode spectrum in the limiting case of the thin shell is
significantly different from that of the filled sphere, which
should be testable in the Cold Atom Laboratory (CAL) trap, (ii) a
dip in the frequency of breathing-type or radial collective modes
(having nodes purely along the radial direction) accompanies the
hollowing transition when an inner boundary first appears, (iii) a
reconfiguration of the surface (high-angular-momentum) collective
modes occurs at the hollowing transition, due to the appearance of
the new surface. By employing a family of trapping potentials, we
show that the dip feature is universal to the hollowing
transition, but its sharpness depends on the details of how
central condensate density depletes. Additionally, we identify
features in the surface mode spectra that are specific to the
hollowing in a bubble trap that would be absent for other trap
geometries (such as the degeneracy in frequencies of modes
localized at the inner and outer condensate shell surfaces,
respectively). As the emergence of an additional boundary over the
course of the hollowing-out deformation renders the filled sphere
and the hollow shell condensates topologically distinct, our
discussion identifies concrete, experimentally testable, features
of a real-space topological transition.

Turning to the experimental realization of a shell condensate,
this requires a combination of factors, all expected to be
achieved in the near future. The bubble trap, first envisioned by
Zobay and Garraway~\cite{Zobay2001}, relies on radio-frequency
(rf) dressing. Recently, the technique has been successfully used
for double-well interferometry, ring-trap, and bubble-trap studies
in one, two, and three dimensions,
respectively~\cite{Schumm2005,Hofferberth2006,
Lesanovsky2006,Jo2007,Fortagh2007,Sinuco-Leon2015,Colombe2004,Merloti2013}.
The 3D system can suffer sag due to the effect of gravity on the
system. While previous experimental work with this geometry has so
far been limited to the disk-shaped BEC produced under large
gravitational sag~\cite{Colombe2004,Merloti2013}, a fully covered
shell-shaped condensate may be produced in the presence of gravity
if particular experimental parameters can be achieved.
Alternately, the effects of gravity can be lessened by performing
the experiment in microgravity. Two experimental microgravity
facilities currently exist: the Zentrum fur Angewandte
Raumfahrttechnologie Und Mikrogravitation (ZARM) drop
tower~\cite{ZARM} in Bremen, Germany and NASA's CAL aboard the
International Space Station~\cite{CAL}. A series of experiments
employing a bubble trap in the latter setting is expected to
investigate the physics of closed BEC shells.

The role of gravity in the realization and behavior of BEC shells
is thus significant. In addition to the gravity-free collective
mode analysis, we therefore include the effects of gravity in two
ways. First, we estimate the strength of gravity required to alter
the BEC shell structure and show that typical strengths on Earth
far surpass this limit. We also provide density profiles of shells
in the presence of a gravitational field and show the sag effect.
The effect, as a function of field strength, opens up the shell at
a pole, altering the topology, and progresses to flatten the
opened shell. Second, assuming microgravity conditions, we
perturbatively analyze the effects of weak gravity on collective
modes.  As might be expected, the field breaks spherical symmetry,
and modes differing by a unit of angular momentum in the
spherically symmetric case become coupled.

Our comprehensive study of the evolution of a BEC from a filled
sphere to a thin shell serves multiple purposes. It introduces a
balance of analytic and numerical techniques appropriate for
studying condensate equilibrium profiles and collective mode
structures in hollow geometries. It presents concrete results for
detecting a hollowing transition in such systems by pinpointing
signatures in the BEC's collective mode spectrum associated with
the topological change of acquiring a new inner boundary.
Appropriate to realistic settings, it assesses the effects of
gravity and provides experimental estimates.  We begin in
Sec.~\ref{sec:II} with a description of equilibrium properties of
a BEC system in trapping potentials capable of evolving from a
filled sphere to a thin shell as a function of an experimentally
accessible tuning parameter. We incorporate a combination of
techniques that best capture the associated BEC density profiles.
We then introduce two techniques in Sec.~\ref{sec:III} for
studying dynamics using hydrodynamics in one case and numerical
evolution following a sudden trap change in the other. In
Sec.~\ref{sec:IV}, we discuss the collective mode structures for
the limiting cases of the well-known filled sphere as well as the
thin shell. In Sec.~\ref{sec:V}, we perform a thorough analysis of
the collective mode structure evolution between the limiting
cases, first analyzing radially symmetric modes and then exploring
higher angular momentum modes including surface modes. In
Sec.~\ref{sec:VI}, we show that features in the evolution of the
collective mode spectra can be attributed to the hollowing of the
condensate density at the center of the system. In
Sec.~\ref{sec:VII}, we turn to the presence of gravity, obtaining
bounds for when the shell structure can be preserved in spite of a
gravitational sag We also perform a perturbative treatment of
gravitational effects on the collective mode structure. We then
end our exposition with considerations for realistic experimental
settings in Sec.~\ref{sec:VIII} and a conclusion in
Sec.~\ref{sec:IX}.

\section{Equilibrium Profiles}\label{sec:II}

Here, we establish the equilibrium properties of hollow shell BECs
upon which we build our collective mode description. We consider
the limiting cases of a harmonic trap for the filled sphere
geometry and a radially shifted harmonic trap for the thin shell
condensate. A ``bubble trap" proposed in the
literature~\cite{Zobay2001} and related to the experimental CAL
trap can be tuned between these two limiting cases. We then
consider a generalized trapping potential which can tune the
condensate equilibrium density in the center of the system as it
hollows, thereby allowing us to isolate the effect of density in
the creation of an inner boundary. We analyze the equilibrium
condensate density in these traps and note that the Thomas-Fermi
approximation (which neglects the kinetic energy of the
condensate) is a good approximation for many analyses, and
pinpoints hollowing-out features by modeling sharp boundaries for
the condensate. Since accounting for realistic condensate profiles
and soft boundaries requires going beyond the Thomas-Fermi
approximation, we also employ numerical methods for determining
ground-state densities.

We describe the condensate wave function $\psi ({\bf{r}},t)$
within the standard time-dependent Gross-Pitaevskii (GP) equation,
given by
\begin{eqnarray}\label{eq:GPeqn}
i\hbar {\partial _t}\psi ({\bf{r}},t) = \left[ { - \frac{{{\hbar
^2}}}{{2m}}{\nabla ^2} + V({\bf{r}}) + U_0{{\left| {\psi
({\bf{r}},t)} \right|}^2}} \right]\psi ({\bf{r}},t), \nonumber\\
\label{eq:GP}
\end{eqnarray}
where $m$ is the particle mass, $V$ is the trapping potential, and
$U_0=4 \pi \hbar^2 a_s /m$ is the interaction strength
(proportional to the two-body scattering length $a_s$)
\cite{Pethick2008}. In considering equilibrium properties, we
employ the time-independent version of the GP equation obtained by
assuming a condensate wavefunction of the stationary form $\psi
({\bf{r}},t) = \psi ({\bf{r}})\exp(-i\mu t/\hbar)$ , where $\mu$
is the chemical potential of the system. The equilibrium
condensate density is then given by $n_{\rm{eq}}(\mathbf{r}) =
|\psi ({\bf{r}})|^2$.

\subsection{Trapping potentials}\label{sec:IIA}

We begin by defining trapping potentials that host the limiting
cases of the filled spherical condensate and the thin condensate
shell, and then discuss two possible trapping potentials for
capturing the physics of the intermediate shell regime.
Equilibrium density distributions of condensates are primarily
determined by the trapping potential, with details such as their
widths and maximum radii being influenced by the strength of
interactions between the atoms. In the equations below, we adopt
dimensionless length units, rescaled by an oscillator length $S_l
= \sqrt{\hbar/(2m\omega)}$, where $\omega$ is a relevant
frequency, for instance, of the bare harmonic confining trap prior
to rf dressing.

As the simplest and best understood case, we consider the
spherically symmetric harmonic trap, which produces a fully filled
spherical condensate. The associated trapping potential takes the
form
\begin{equation}\label{eq:harmonic}
V_{\rm{0}}(r)=\frac{1}{2}m\omega_0^2 S_l^2 r^2,
\end{equation}
where $\omega_0$ is the single-particle frequency of small
oscillations and $r$ is the dimensionless radial distance from the
origin (spherical center).

In the opposite limit of a very thin spherical condensate shell, a
simple trapping potential that would produce this shape is a
radially shifted harmonic trap of the form,
\begin{eqnarray}\label{eq:Vsh}
V_{\rm{sh}}(r)=\frac{1}{2}m\omega_{\rm{sh}}^2 S_l^2 (r-r_0)^2,
\end{eqnarray}
where the location of potential minimum $r_0$ is nonzero, and
$\omega_{\rm{sh}}$ is the frequency of single-particle
oscillations near this radius. Several salient features of the
equilibrium density and collective mode structure of the thin
shell limit are well captured by analyzing condensates in this
radially shifted potential. In addition, this potential is a good
approximation for any trapping potential with a radially shifted
minimum in the thin shell limit.

However, the radially shifted harmonic potential is unphysical for
situations where the condensate density is finite close to the
trap center; the slope of the potential is discontinuous at $r=0$.
For a more realistic trap producing a hollow condensate, we
consider a ``bubble
trap"~\cite{Zobay2001,Colombe2004,Merloti2013}, which has been
recognized as a good candidate.
 Such a trap is achievable in the cold atomic setting
by employing time-dependent, radio frequency induced adiabatic
potentials within a conventional magnetic trapping geometry.
Its form is given by
\begin{equation}\label{eq:bubble}
V_{\rm{bubble}}=m\omega_0^2 S_l^2
\sqrt{(r^2-\Delta)^2/4+\Omega^2},
\end{equation}
where $\Delta$ and $\Omega$ are the effective (dimensionless)
detuning between the applied rf-field and the energy states used
to prepare the condensate and the Rabi coupling between these
states, respectively. Note that the minimum of this potential is
found at $r=\sqrt{\Delta}$ and the frequency of single-particle
small oscillations around this minimum is $\sqrt{\Delta/\Omega}
\omega_0$.

The parameters $\Delta$ and $\Omega$ together allow for tuning
between the filled condensate and the thin shell. When
$\Delta=\Omega=0$, the bubble-trap potential reduces to a harmonic
trap with frequency $\omega_0$. For large $\Delta$, it is
approximated near its minimum by a radially shifted harmonic trap
[Eq.~(\ref{eq:Vsh})] with frequency $\omega_{\rm{sh}} =
\sqrt{\Delta/\Omega} \omega_0$. Slowly increasing or decreasing
the trap parameter $\Delta$ results in a continuous deformation
between the two limiting geometries of a filled sphere and a thin
spherical shell.

While the change from a filled to a hollow system is topological
in nature (an inner surface is created), because condensates have
continuous density profiles, the hollowing in a real system will
be gradual. The density at the center  becomes smaller until it is
effectively zero. One of the main questions addressed in this work
is the following: What are the signatures of this transition in
the collective mode spectrum of the system?

In order to tune the detailed behavior of the condensate density
during the hollowing transition, we also consider the following
general radially shifted trapping potential
\begin{eqnarray}\label{eq:gt_potential}
V_{\rm{gt}}(r) = \frac{1}{2}m\omega^2_{\rm{gt}} R^2 S_l^2 {\left[ {{{\left( {\frac{r}{R}}
\right)}^\alpha } - \gamma } \right]^2},
\end{eqnarray}
where $R$ represents the (dimensionless) characteristic size of
the system (or exactly the outer sharp boundary under the
Thomas-Fermi approximation). Here, $0 \le \gamma \le 1$ is a
dimensionless parameter that tunes the radially shifted trap
minimum for realizing the evolution between filled sphere and thin
shell, and $\alpha$ determines the polynomial growth of condensate
density in the radial direction from the boundary. Note that the
radially shifted harmonic trap of Eq.~(\ref{eq:Vsh}) is a special
case of the general trap having $\alpha=1$. In Sec.~\ref{sec:VI},
we will use this general potential to understand the universal
features of the collective-mode spectrum that occur as the system
becomes hollow in the center.

\subsection{Evolution of equilibrium density from filled sphere to thin shell}\label{sec:IIB}

Having enumerated the trapping potentials used in our studies, we
next find the equilibrium density profiles of condensates confined
to these traps.

\subsubsection{Thomas-Fermi approximation}

In the limit of strong interactions the Thomas-Fermi
approximation, in which the kinetic energy term in
Eq.~(\ref{eq:GP}) is disregarded, is commonly used. In this
approximation, the equilibrium density profile is given by
\begin{eqnarray}
n_{\rm{eq}}(\mathbf{r})=\frac{V(\mathbf{R})-V(\mathbf{r})}{U},\label{eq:neq}
\end{eqnarray}
where $R$ is the outer radius of the condensate, $U=U_0 S_l^{-3}$,
and the chemical potential is equal to $\mu=V(R)$. The strong
interaction limit is defined by  $N a_s/S_l \gg 1$ for $N$
particles, where $S_l$ is the characteristic length of the
trapping potential defined in Sec.~\ref{sec:IIA}.

Within this approximation, the condensate density
$n_{\rm{eq}}(\mathbf{r})$ and boundary $n_{\rm{eq}}(\mathbf{R})=0$
are determined by the trap geometry and the total number of
particles $N = \int {{n_{{\rm{eq}}}}({\bf{r}})d{\bf{r}}}$. This
approximation can be used to show that a BEC in the bubble trap
with a slowly changing detuning $\Delta$ transitions from a filled
sphere to a hollow shell when $\Delta = R^2/2$ (where we have set
$\Delta=\Omega$ for convenience).

\begin{figure}[t]
\centering
\includegraphics[width=6.5cm]{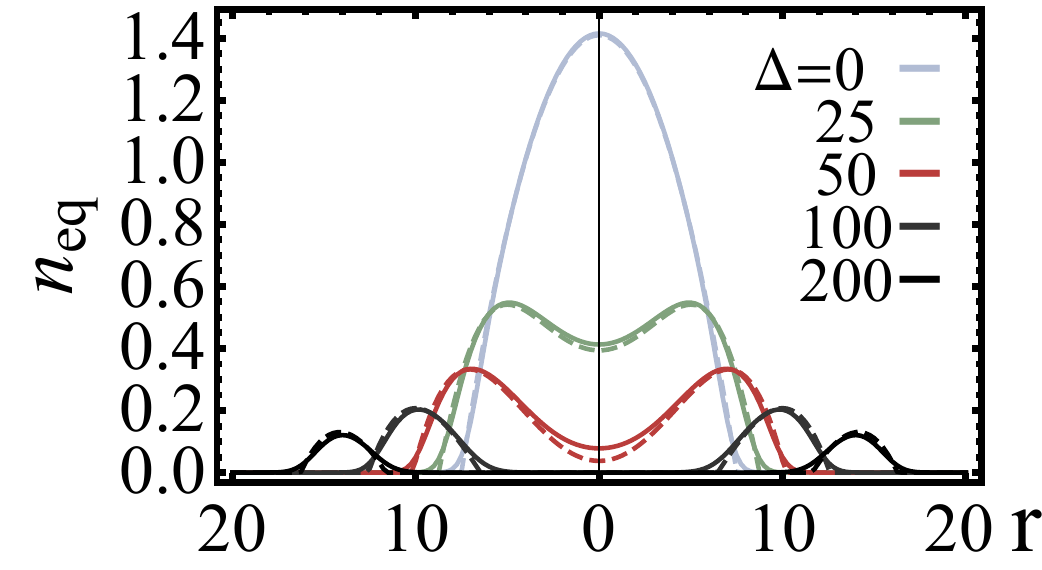}
\caption{(Color online) Ground-state density profiles
$n_{\rm{eq}}(r)$ in the bubble trap for $\Delta = 0$ (lighter blue
curves), 25 (light green), 50 (dark red), 100 (darker gray), and
200 (black). The various values of $\Delta$ show the evolution
from the spherical (filled) condensate in a harmonic trap at
$\Delta=0$ to a thin-shell condensate for large $\Delta$ (the 3D
visualization can be found in the top row of Fig.~\ref{fig:1}).
The solid (dashed) curves are obtained from the numerical solution
of the GP equation (the Thomas-Fermi approximation). Here, we use
$\Delta/\Omega=1$ and an interaction constant $u=10,000$.}
        \label{fig:2}
\end{figure}

\subsubsection{Beyond Thomas-Fermi}
A more accurate numerical
solution for the ground state of the GP equation (for any value of
interaction strength) can be found using an imaginary-time
algorithm~\cite{ChiofaloI2000}. In Fig.~\ref{fig:2}, we show the
evolution of the numerical ground state in the bubble trap as the
detuning $\Delta$ is varied from $0$ to $200$ (corresponding to
evolution from a filled, harmonically trapped, sphere to a thin
spherical shell) with $\Delta/\Omega=1$ fixed. The equilibrium
density given by the Thomas-Fermi approximation is shown for
comparison. While for the most part the density profiles match
each other, those found through the exact numerical calculation
show a more realistic, smooth decrease in density at the edges of
the condensate.

We note that for intermediate interaction strength $U$, the
relative interaction energy (as measured by the ratio of $
\frac{U_0}{2} \int |\psi|^4 d\mathbf{r}$ to the total energy) can
decrease considerably as one moves from a filled sphere to a thin
shell when the number of trapped particles $N$ is kept fixed. This
is due to the large decrease in maximum particle density
$n=|\psi|^2$ as the shell becomes thin. In order to compare with
results of the Thomas-Fermi approximation in the thin shell limit,
we have used a relatively large value of dimensionless interaction
strength $u=8\pi N a_s/S_l=10,000$ in our numerics, where $S_l$ is
the oscillator length for the harmonic trap when $\Delta=0$. For
$^{87}$Rb and a bare trap frequency of 10--100 Hz, for instance,
this corresponds to $N=2\times 10^5$ to $5\times 10^4$ atoms.

Taking into account the analyses and arguments above, in what
follows, we either employ the Thomas-Fermi approximation or the
numerical ground state, as appropriate.

\section{Methods and approaches for dynamics of collective modes}\label{sec:III}

Having discussed the equilibrium behavior of a trapped condensate,
in this section we consider its dynamics. In order to study the
collective motions of spherically symmetric BECs we present four
complementary methods appropriate in different regimes: (i) The
first method uses a hydrodynamic approach combined with
Thomas-Fermi equilibrium density profiles, which leads to a
differential eigenvalue problem for all possible collective
motions of small density deviations of a given condensate
geometry. This method provides a good approximation for collective
mode spectra in any spherically symmetric trap. (ii) Second, we
employ a fully numerical approach that mimics the experimental
excitation of collective oscillations through sudden changes in
trap potential. While more realistic, this method can only capture
low-lying collective modes due to finite resolution. (iii) A
hybrid method uses the hydrodynamic equations, but applied to the
numerical ground-state density and has the advantages of the
simpler Thomas-Fermi hydrodynamic approach, but with realistic
treatment of the condensate boundaries. (iv) A fourth method
focuses on surface modes localized to the boundary of the system
by linearizing the hydrodynamic eigenproblem close to the
boundaries of the condensate.

\subsection{Hydrodynamic treatment}\label{sec:IIIA}
Here, we describe the condensate's collective dynamics in terms of
hydrodynamic density oscillations. The small density fluctuation
$\delta n(\mathbf{r},t)$ out of the equilibrium condensate density
$n_{\rm{eq}}(\mathbf{r})$ is described by the standard
hydrodynamic equation of motion (see details in
Refs.~\cite{Stringari1996,Pethick2008}),
\begin{eqnarray}
mS_l^2\partial _t^2\delta n = U\nabla  \cdot \left(
{{n_{{\rm{eq}}}}\nabla \delta n} \right) + O(\delta
{n^2},\partial^3 \delta n,\partial
^2n_{{\rm{eq}}}).\label{eq:delta_n}
\end{eqnarray}
We will use this expression for specific trap geometries under a
number of assumptions about the nature of the collective modes.
First, the nonlinear terms of $\delta n$ are negligible when we
assume that deviations away from the equilibrium density are
small, i.e., the amplitude of the collective mode oscillations is
small. Furthermore, the higher derivatives of $\delta n$ are
negligible if the oscillations in the condensate density are
smooth, i.e., $\delta n$ varies only on length scales much larger
than the local coherence length $a_d\sqrt{a_d/a_s}$ where $a_d$ is
the interparticle spacing~\cite{Pethick2008}. The second
derivative of $n_{\rm{eq}}$, associated with the kinetic energy
arising from the density variation due to the trap, is also
negligible in the strong interaction (Thomas-Fermi) limit.

Assuming normal-mode oscillations $\delta n(\textbf{r},t)=\delta
n(\textbf{r})e^{i\omega t}$, Eq.~(\ref{eq:delta_n}) takes the
form,
\begin{eqnarray}\label{eq:delta_n_no_TF2}
-\frac{mS_l^2}{U}\omega^2\delta n=\nabla n_{ \rm{eq}}\cdot
\nabla\delta n+n_{ \rm{eq}}\nabla^2\delta n.
\end{eqnarray}
The eigenvalues and eigenfunctions satisfying this expression
correspond to collective modes with oscillation frequencies
$\omega$ and profiles $\delta n(\textbf{r})$, respectively.  We
note that this expression shows that the hydrodynamic approach can
be applied for any equilibrium density profile. While for the most
part we will use the density obtained from the Thomas-Fermi
approximation, we will use the numerical ground-state density
found from the imaginary-time algorithm for comparison.

For traps with spherical symmetry, we can decompose $\delta
n({\bf{r}})=D(r)Y_{\ell m_{\ell}}(\theta, \varphi)$, where
$Y_{\ell m_{\ell}}$ are the usual spherical harmonics and find
that Eq.~(\ref{eq:delta_n_no_TF2}) reduces to
\begin{eqnarray}\label{eq:eigenproblem_noTF}
\frac{mS_l^2}{U}{\omega ^2 r^2}D = - \frac{d}{dr} \left( r^2
n_{\rm{eq}} \frac{d D}{dr} \right )+\ell(\ell+1)n_{\rm{eq}}D.
\end{eqnarray}
Using the Thomas-Fermi equilibrium density profile of
Eq.~(\ref{eq:neq}) the eigenproblem becomes
\begin{eqnarray}\label{eq:eigenproblem}
mS_l^2{\omega ^2}D &=& \frac{{dV}}{{dr}}\frac{{dD}}{{dr}} - \left[
{V(R) - V(r)} \right] \nonumber\\ &&\times \left[
{\frac{{d{D^2}}}{{d{r^2}}} + \frac{2}{r}\frac{{dD}}{{dr}} -
\frac{{\ell(\ell + 1)}}{{{r^2}}}D} \right].
\end{eqnarray}
Equations (\ref{eq:eigenproblem_noTF}) and (\ref{eq:eigenproblem})
have analytic solutions in two limiting cases of filled sphere and
thin shell, which are studied in Sec.~\ref{sec:IV}, or can be
analyzed with perturbation theory, as for the finite thin-shell
case in Sec.~\ref{sec:IVB} or for the gravity effects studied in
Sec.~\ref{sec:VIIB}. In general, our eigenproblem is of the
Sturm--Liouville form and can hence be treated with a
finite-difference method, whose technical details are presented in
Appendix~\ref{app:Finite_diference}. Our hydrodynamic results for
the evolution between filled sphere and hollow shell in
Secs.~\ref{sec:V} and \ref{sec:VI} are obtained from the
finite-difference method.

\subsubsection{Surface modes}
Here, we highlight the case of
collective modes localized near the edges of the condensate---its
surface modes. Following the earlier work in
Ref.~\cite{AlKhawaja1999}, we note that in order to identify
surface modes it is appropriate to expand the Thomas-Fermi
equilibrium density profile about $\mathbf{r}=\mathbf{r}_b$ for
$V(\mathbf{r}_b)=\mu$, the position of the condensate boundary.
Consequently, the Thomas-Fermi equilibrium density profile can be
expressed as
\begin{eqnarray}
n_{\rm{eq}}(\mathbf{r})=\frac{\mathbf{F}\cdot
(\mathbf{r}-\mathbf{r}_b)}{U}, \label{eq:neq_surface}
\end{eqnarray}
where $\mathbf{F}=-\nabla V(\mathbf{r}_b)$.

Denoting the direction of $\nabla V$ by $x$ and the position of
the boundary (which is a 2D equipotential surface) by $x=x_b$, we
proceed to estimate the size of the region near the boundary
within which this treatment of the condensate equilibrium density
is reliable. Recalling that the Thomas-Fermi approximation follows
from neglecting the contribution of the kinetic energy in the GP
equation, Eq.~(\ref{eq:GPeqn}), we estimate that kinetic energy
dominates for $x_b-x\leq \delta_{\rm{sm}}$ where
\begin{eqnarray}\label{eq:KEdelta}
  \delta_{\rm{sm}}=\left(\frac{\hbar}{2m|\mathbf{F}|}\right)^{1/3}.
\end{eqnarray}
If the trapping potential varies slowly on the length scale
$\delta_{\rm{sm}}$, it is then a good approximation to expand the
potential and the Thomas-Fermi equilibrium density about $r=r_b$
as suggested above~\cite{Pethick2008}.

We next solve the hydrodynamic equations for the linearized
potential $V(x)=Fx$ with $x$ defined so that it vanishes at $r_b$.
First, we note that in the $y$ and $z$ directions, defined
relative to $x$, there is translational invariance so that a
density oscillation $\delta n=\delta n(x,y,z)e^{-i\omega t}$ must
correspond to a plane wave in these two variables. As in
Ref.~\cite{AlKhawaja1999}, we chose the $z$-direction as the
direction of propagation of this collective mode and denote its
wave number by $q$. Consequently, the collective motion only has
$x$ and $z$ dependences and we can rewrite the hydrodynamic
equation as
\begin{eqnarray}\label{eq:hydrox}
mS_l^2\omega^2\delta n(x,z) &=& F\frac{\partial\delta n(x,z)}{\partial x}+Fx\left[\frac{\partial^2\delta n(x,z)}{\partial x^2} \right. \nonumber\\
&&\left.+\frac{\partial^2\delta n(x,z)}{\partial z^2}\right].
\end{eqnarray}
Solutions to this expression can take the form
\begin{eqnarray}\label{eq:ansatz}
\delta n(x,z)=f(qx)e^{qx+iqz}
\end{eqnarray}
for some smooth function $f(qx)$. We note that in this ansatz, the
exponential factor in the variable $x$ corresponds to an
exponential decay away from the condensate boundary. We obtain the
frequencies for modes described by Eq.~(\ref{eq:ansatz}) as
\begin{eqnarray}\label{eq:dispersion}
mS_l^2\omega_{\rm{in,out}}^2=(1+2\nu_{\rm{in,out}})F_{\rm{in,out}}q_{\rm{in,out}},
\end{eqnarray}
while the modes themselves are of the form
\begin{eqnarray*}
\delta
n(x,y,z,t)&=&C_{\nu_{\rm{in,out}}}L_{\nu_{\rm{in,out}}}(-2q_{\rm{in,out}}x)
\nonumber\\
&& \times
e^{q_{\rm{in,out}}x+iq_{\rm{in,out}}z-i\omega_{\rm{in,out}} t},
\end{eqnarray*}
where $C_{\nu_{\rm{in,out}}}$ is the overall magnitude and  $L_{\nu_{\rm{in,out}}}$
are Laguerre polynomials. Here, indices
$\nu_{\rm{in,out}}$ count the number of radial nodes of a
collective mode confined to a particular boundary surface of the
condensate.

The specific behavior of surface modes, depending on the specific
physical topology of the condensate, is further discussed in
Sec.~\ref{sec:IVC} for the limiting shell case and in
Secs.~\ref{sec:VB} and \ref{sec:VIC} for the evolution between
filled sphere and hollow shell in the bubble and general traps,
respectively.

\subsection{Sudden quench numerics}\label{sec:IIIB}
While the hydrodynamic treatment described above is adequate for
obtaining the collective mode spectrum, our numerical simulations
serve as a good complement. They even capture physics beyond the
hydrodynamic regime for the bubble trap geometry, which can
access the filled sphere and thin shell limits as well as the
evolution between them. Taking inspiration from experimental
methods for exciting collective modes of trapped
BECs~\cite{Mewes1996}, we theoretically probe the excitations of
the system by direct simulation of the GP equation after a small
and sudden change in the trap potential takes the system slightly
out of equilibrium.

After first finding the numerical ground state of the GP equation
describing a BEC confined by the bubble trap for one set of values
of trap parameters $\Delta_0$ and $\Omega_0$, we then time-evolve
this initial state using the GP equation with different values of
trap parameters $\Delta$ and $\Omega$. This results in
time-oscillating features in the time evolved condensate
wavefunction $\psi ({\bf{r}},t)$. The frequencies of these
oscillations can be extracted by using a fast-Fourier transform in
the time domain. For small changes ($\Delta_0-\Delta \ll \Delta_0$
and likewise for $\Omega$) these simulations should probe the
linear response and low-lying modes of the system.

In practice, we use the explicit time-marching method of
Ref.~\cite{Cerimele2000} to time evolve the initial numerical
ground state (found in Sec.~\ref{sec:IIB}) for two types of
quenches. In one case, we let $\Delta/\Omega\neq
\Delta_0/\Omega_0$ change while keeping $\Delta=\Delta_0$ fixed,
which fixes the mean radius of the shell but allows the tightness
of the confinement to change. For a hollow shell, this quench is
expected to primarily excite modes in which the thickness of the
shell oscillates. Alternatively, we change $\Delta\neq \Delta_0$
while keeping $\Delta/\Omega=\Delta_0/\Omega_0$ fixed so that the
tightness of the trap is constant while the mean radius of the
shell varies. This method is found to more equally excite modes
with both even and odd values of $\nu$ (which denotes the number
of radial nodes in the collective mode). The two methods agree
when the same mode frequencies are resolvable. Because these
quenches both preserve spherical symmetry, they only probe
spherically symmetric ( $\ell=0$) modes of the system. In
principle, appropriate quenches could be designed to probe
nonsymmetric states as well (such as an offset of the center of
the trap in the $x$ direction, resulting in excitation of the
center-of-mass modes). In what follows, we concentrate on
numerical resolution of the spherically symmetric modes only.

\section{Collective modes: limiting cases}\label{sec:IV}

In this section, we apply the hydrodynamic approach of
Sec.~\ref{sec:IIIA} to a filled spherical condensate and a thin
condensate shell. In this detailed treatment, we establish the
full frequency spectrum and associated mode behaviors of these two
limiting cases of our general hollowing system. Because of the
spherical symmetry of the system, the collective modes can be
characterized by two quantum numbers: one for the number of radial
nodes, $\nu$, and one for the orbital angular momentum denoted by
the index $\ell$. We find that the presence of an inner boundary
in condensate shells produces features that are not found in fully
filled spherical BECs even for the same quantum numbers. These
results not only provide a check for the full evolution between
the limits when analyzing the bubble trap and other cases in
subsequent sections, they also provide a better understanding of
the intermediate crossover regime and corroborate predictions for
this regime.

\subsection{Filled sphere: Quadratic potential}\label{sec:IVA}

We first recapitulate the results of hydrodynamic approach for the
well-understood case of a condensate in a spherically symmetric
harmonic trap, Eq.~(\ref{eq:harmonic}), $V_{\rm{0}}=\frac{1}{2}m
{\omega_0}^2 S_l^2 r^2$. We assume the strong interaction regime
where the Thomas-Fermi approximation is applicable.

Solving Eq.~(\ref{eq:eigenproblem}), one obtains
\cite{Stringari1996,Pethick2008}
\begin{equation}\label{eq:spheremodes}
\omega_{\nu,\ell}^{\rm{sp}}=\omega_0\sqrt{\ell+3\nu+2\nu \ell +2
\nu^2},
\end{equation}
where $\nu$ is a radial index as above and $\ell$ is an index of
the orbital angular momentum. We see from
Eq.~(\ref{eq:spheremodes}) that in addition to the trivial mode
$\omega_{0,0}^{\rm{sp}}=0$, which corresponds to a uniform density
deviation but does not physically exist in real systems, any
nonzero mode has frequency $\omega_{\nu,\ell}^{\rm{sp}} \ge
\omega_0$. A schematic density deviation profile for $(\nu,\ell)=
(1,0)$ [$(1,1)$] is presented in the leftmost panel of the second
(third) row of Fig.~\ref{fig:1}. We see that $\nu$ and $\ell$
count the nodes, at which the density deviation vanishes, in
radial and angular directions, respectively. The $\ell=0$ modes
exhibit only the radial expansion and contraction (spherically
symmetric) and are referred as breathing modes. The $\ell=1$ modes
exhibit a center-of-mass oscillation between the southern and
northern hemispheres and are referred as sloshing or dipole modes.
The breathing, sloshing, and quadrupole ($\ell=2$) modes have been
experimentally
observed~\cite{Jin1996,Mewes1996,Lobser2015,Straatsma2016} to be
in good agreement with the theoretical predictions.

\subsection{Thin shell: Radially shifted quadratic potential}\label{sec:IVB}
As discussed in Sec.~\ref{sec:II}, the shell-shaped condensate can
be realized by trapping in a radially shifted harmonic potential,
Eq.~(\ref{eq:Vsh}),
\begin{equation*}
V_{\rm{sh}}(\mathbf{r})=\frac{1}{2}m \omega_{\rm{sh}}^2S_l^2
(r-r_0)^2.
\end{equation*}
The potential minimum (or condensate density maximum) appears at
$r=r_0$. Under the Thomas-Fermi approximation, given the shell's
outer radius $R=R_{\rm{out}}=r_0+\delta$ and $\delta < r_0$,
Eq.~(\ref{eq:neq}) yields the inner radius
$R_{\rm{in}}=r_0-\delta$. Here, the condensate shell has two
distinct radii, $R_{\rm{in}}$ and $R_{\rm{out}}$, in contrast to a
single outer radius $R$ for the fully filled spherical condensate.
Evaluating the integral $N = \int
{{n_{{\rm{eq}}}}({\bf{r}})d{\bf{r}}}$, we find the relationship
between the total number of particles, $N$, and the shell's
dimensionless thickness, $2 \delta$:
\begin{equation}
N =\frac{{8\pi m\omega _{{\rm{sh}}}^2S_l^2}}{{3U}}{\delta ^3}(r_0^2
+ \frac{{{\delta ^2}}}{5}).
\end{equation}
If $\delta \ll r_0$, this reduces to
\begin{equation} \label{eq:shellthick}
2 \delta = {\left( {\frac{{3UN}}{{\pi m\omega
_{{\rm{sh}}}^2 S_l^2 r_0^2}}} \right)^{1/3}}.
\end{equation}
We define the ``thin shell limit" by $c \equiv r_0/\delta \gg 1$
and solve Eq.~(\ref{eq:eigenproblem}) for the collective modes in
this limit.

We let $\eta=\frac{r-r_0}{\delta}$ (so $-1\leq{\eta}\leq{1}$),
$\lambda=\left( \omega/\omega_{\rm{sh}} \right)^2$, as well as
$y(\eta)/(\eta+c)=D(r)$ and express Eq.~(\ref{eq:eigenproblem}) as
\begin{eqnarray}
&&\Big [(1-\eta^2)\frac{d^2}{d\eta^2}-2\eta\frac{d}{d\eta}+2{\lambda}\nonumber\\
&&+\frac{2\eta}{\eta+c}-\ell(\ell+1)\frac{1-\eta^2}{(\eta+c)^2} \Big]y=0.
 \label{eq:dfshell}
\end{eqnarray}
In the limit of a very thin shell ($c \to \infty$), we can ignore the last two terms in Eq.~(\ref{eq:dfshell}). The solutions are Legendre polynomials
$y_\nu(\eta)=\sqrt{\frac{2\nu+1}{2}}P_\nu(\eta)$ having eigenfrequencies
\begin{equation}
\omega_{\nu,\ell}^{\rm{sh}}=\omega_{\rm{sh}}\sqrt{\nu(\nu+1)/2}.
\label{eq:thin_shell_modes}
\end{equation}
Here, similar to the filled sphere case, the thin shell case has
nonzero breathing mode frequencies higher than the characteristic
trapping frequency. But in contrast to the filled case, all $\ell$
modes are nearly degenerate compared with the radial energy scale.
The rightmost panel of second (third) row of Fig.~\ref{fig:1}
shows a schematic density deviation profile for $(\nu,\ell)=
(1,0)$ [$(1,1)$], from which we see  nodal structures similar to
the filled sphere case. For a very thin shell, however, the energy
associated with a collective mode is largely unaffected by the
number of angular nodes.  In other words, the radial behavior of
the collective mode determines its eigenfrequency. We note that
this degeneracy of $\ell$ modes has previously been
shown~\cite{Stringari2006} to characterize the collective modes of
thin ring-shaped BECs as well. Additionally, we note that, unlike
the indices $\nu_{\rm{in,out}}$ of Eq.~(\ref{eq:dispersion}),
$\nu$ denotes the number of radial zeros of a collective mode
spanning the full extent of the condensate shell.

For a slightly thicker shell in which $c$ deviates from the
infinite limit, we calculate the correction to
$\omega_{\nu,\ell}^{\rm{sh}}$ by treating the last two terms in
Eq.~(\ref{eq:dfshell}) perturbatively. We obtain
\begin{eqnarray}\label{eq:thinshellpert}
\left(\frac{\omega_{\nu,\ell}^{\rm{sh}}}{\omega_{\rm{sh}}}
\right)^2&&=\frac{\nu(\nu+1)}{2}\left[1+\frac{4c^{-2}}{(2\nu-1)(2\nu+3)}\right]\nonumber\\
&&+
\frac{c^{-2}\ell(\ell+1)}{4}\left[1-\frac{1}{(2\nu-1)(2\nu+3)}\right].
    \label{eq:shell_spec}
\end{eqnarray}

The leading corrections to the frequency are of the order
${O}(c^{-2})$. In a shell with a large but finite $c$, the
frequency spectrum of the lowest lying collective modes (low $\nu$
and $\ell$ indices) has the form of bands corresponding to
different $\nu$ separated by ${O}(\omega_{\rm{sh}})$, with each
band having fine levels corresponding to different $\ell$
separated by ${O}(c^{-2} \omega_{\rm{sh}} )$. As a result, the two
energy scales for radial and angular motions are well separated in
a thin shell. As the thickness of the shell increases, the effects
of the angular oscillations on its collective mode frequency and
energy become more prominent.

In the $\nu=0$ case, we obtain purely angular collective modes
\begin{eqnarray}\label{eq:angularmodes}
\omega_{0,\ell}^{\rm{sh}}=c^{-1}\omega_{\rm{sh}}\sqrt{\ell(\ell+1)/3}.
\end{eqnarray}
Accordingly, we expect the collective modes of a very thin shell
that do not have any radial nodes to correspond to very low (but
still nonzero) frequencies. Such low-frequency excitations
(compared with the trap frequency) do not exist in the
filled-sphere condensate.

Additionally, we note that the presence of an inner boundary for a
shell condensate can have important effects on its collective
modes. In the limit of large $\ell$ ($\gg c$), the term
$\ell(\ell+1)(1-\eta^2)/(\eta+c)^2$ in Eq.~(\ref{eq:dfshell})
dominates. This term behaves as a potential barrier between the
inner and outer boundaries $\eta = \pm 1$, respectively, and
favors low-frequency modes localizing on either the inner or outer
shell surface (potential minimum). This is different from the
filled-sphere condensate, which does not have an inner boundary.
We proceed to discuss these surface modes below, and later in
Sec.~\ref{sec:VB}, we show how this effect causes a sudden change in
the spectrum of large $\ell$ modes as a bubble-trap system evolves
from a sphere to a shell.

\subsection{Surface modes}\label{sec:IVC}
Noting that the most striking difference between a fully filled
spherical BEC and a hollow condensate shell is the presence of an
additional, inner boundary for the latter, we employ the
techniques presented in Sec.~\ref{sec:IIIA} in order to study the
collective modes localized at condensate boundaries.

For the collective modes localized at the outer edge of the fully
filled spherical BEC it is known that~\cite{AlKhawaja1999}
\begin{eqnarray}\label{eq:spsurface}
 \omega^{\rm{sp}}_{\nu,\ell}=\omega_0\sqrt{\ell(2\nu+1)}.
\end{eqnarray}
In presenting this result,  we note that this expression is
exactly the large $\ell \gg 1$ limit of the collective mode
frequencies given by Eq.~(\ref{eq:spheremodes}).

To study the modes confined to the inner and outer surfaces of the
hollow condensate shell we first consider the radially shifted
quadratic potential  of trapping frequency $\omega_{sh}$ in
Eq.~(\ref{eq:Vsh}) and obtain
\begin{eqnarray}\label{eq:SurfModes}
mS_l^2\omega^2_{\rm{in,out}}=(1+2\nu_{\rm{in,out}})F_{\rm{in,out}}q_{\rm{in,out}},
\end{eqnarray}
or, more precisely,
\begin{eqnarray}\label{eq:VshSurf}
\omega_{\rm{in,out}}^{\rm{sh}}=\omega_{\rm{sh}}\sqrt{\ell(1+2\nu_{\rm{in}})\frac{|R-r_0|}{r_{\rm{in,out}}}},
\end{eqnarray}
where we identify the wave numbers $q=\ell/r_{\rm{in},\rm{out}}$
and calculate $r_{\rm{in}}=2r_0-R$ and
$F_{\rm{in},\rm{out}}=\pm\omega_{\rm{sh}}^2(R-r_0)$. We note that
collective modes on the inner surface have a higher frequency, for
the same number of radial and angular nodes ($\nu$ and $\ell$),
than those on the outer surface. Recalling that the radially
shifted trapping potential captures the salient characteristics of
thin BEC shell dynamics, we note that these expressions are only
representative of condensate behavior for $r_0\approx R$. This
implies that in the thin shell limit, where the areas of the inner
and the outer boundary are comparable, frequencies of surface
modes hosted on either are nearly equivalent as well. At the same
time, we note that this analysis is only applicable to shells of
thickness $\delta > 2\delta_{\rm{sm}}$ as in thinner condensate
shells the kinetic energy near the condensate boundaries cannot be
neglected [recall Eq.~(\ref{eq:KEdelta})] and the surface modes
exhibit significant overlap and thus cannot be treated as fully
confined to either condensate boundary.

This thin-shell result can be compared with the case of a more
general, thicker condensate shell described by the bubble trap of
Eq.~(\ref{eq:bubble}). In this case, the surface modes of the condensate
have frequencies given by
\begin{eqnarray}\label{eq:surfaceomega}
\omega^2_{\mathrm{in,out}}=\frac{\omega_0^2\ell(R^2-\Delta)}{\sqrt{(\Delta-R^2)^2/4+\Omega^2}}(2\nu+1),
\end{eqnarray}
where the wave number associated with each surface mode is
$q_{\rm{in,out}}=\ell/R_{\rm{in,out}}$. In other words, for the
bubble trap, $F_{\rm{in}}q_{\rm{in}}=F_{\rm{out}}q_{\rm{out}}$
leads to a degeneracy in the frequency of surface modes at the
inner and outer surfaces. This degeneracy implies that for a
hollow condensate with two surfaces that are separated by a
substantial thickness, even though the inner surface is smaller in
area, its stiffness is lower and can support more oscillations per
unit distance (since $q_{\mathrm{in}}>q_{\mathrm{out}}$), bringing
the frequency of oscillations with $\nu$ nodes on the two surfaces
into alignment.

In terms of surface modes, we can therefore identify two different
regimes for spherically symmetric BECs: the fully filled BEC
sphere where only the outer condensate boundary is available for
localization of oscillations, and a hollow BEC shell of nontrivial
thickness where the oscillations confined to the inner boundary
attain the same frequencies as those confined to its outer
surface. In the very thin shell limit, these modes overlap, and
their radial nodes do not remain well separated. We note that in
all of these regimes the form of the collective modes is
functionally the same up to factors that explicitly depend on
$r_{\rm{in,out}}$ thus capturing the effects of two boundaries and
finite thickness in the case of hollow condensates. In the
following sections, we discuss the way in which these two regimes
connect as a filled spherical BEC hollows out and deforms into a
thin, hollow shell.

\section{Collective modes: evolution from filled sphere to thin shell }\label{sec:V}

We next examine the evolution of a fully filled spherical
condensate to a hollow, thin shell geometry and its effect on the
system's collective modes. In the extreme limits of the filled
sphere and thin shell, we have given analytic predictions in
Sec.~\ref{sec:IV} above.

In Sec.~\ref{sec:VA}, we show that the collective mode structure
progression for breathing modes from the filled sphere to the thin
shell limits detailed above is characterized by a distinctive
feature---a dip in frequency. The dip occurs at the point in
parameter space when the density at the center of the condensate
first begins to vanish---at the hollowing transition. We find that
at this transition, density deviations for the radial collective
modes localize near the hollowing region. We argue that since the
condensate density in this region is highly reduced, the stiffness
associated with these modes is also lowered at this point,
accounting for the reduced collective mode frequencies.

Density distortions in high angular momentum modes are mainly
confined to the boundary surface of the condensate. The effect of
an emerging surface on the surface mode frequencies is thus
dramatic. In Sec.~\ref{sec:VB}, we detail the evolution of the
surface mode structure through the hollowing transition and find a
distinct rearrangement of the spectrum at the hollowing
transition. We argue that with a new surface present any nodes in
the transverse (radial) direction can be distributed between the
two (inner and outer) boundary surfaces, thus reducing the
energetic cost of hosting these nodes and causing sudden changes
in the mode frequencies.

In this section, we analyze systems in the bubble trap potential
of Eq.~(\ref{eq:bubble}), beginning with a survey of the evolution
of the spherically symmetric mode frequencies and the
corresponding distortions in the condensate density, using the
Thomas-Fermi approximation in the hydrodynamic approach. We then
corroborate and deepen this analysis using the quench numerics
approach of Sec.~\ref{sec:IIIB}. We then analyze finite angular
modes with $\ell \neq 0$, including surface modes. A more focused
analysis is performed in Sec.~\ref{sec:VI} using different
trapping potentials to show that the frequency dip and the surface
mode redistribution are robust for a variety of spherically
symmetric configurations.

\subsection{Evolution of spherically symmetric modes in the bubble trap}\label{sec:VA}

We begin our collective mode analyses by solving the hydrodynamic
differential problem in a bubble trap geometry in the Thomas-Fermi
limit, given by Eq.~(\ref{eq:eigenproblem}), using a
finite-difference method. This calculation is carried out over a
range of mean shell radii $\sqrt{\tilde \Delta} \equiv
\sqrt{\Delta}/R$ thus allowing us to obtain frequencies
corresponding to the same collective mode at various stages of the
evolution between a filled sphere ($\Delta = \tilde \Delta = 0$)
and a very thin shell ($\Delta \approx R^2$ or $\tilde \Delta
\approx 1$). We keep the outer edge of the condensate $R$ fixed
while the total number of atoms is allowed to vary---this
corresponds to working with constant chemical potential $\mu$.
This analysis is expected to capture the physics of a condensate
in the strong interaction limit.

The Thomas-Fermi density profile gives densities corresponding to
a filled sphere at $\tilde \Delta =0$, a filled sphere with depleted central density for $0 <
\tilde \Delta < 0.5$, and a hollow shell for $0.5 < \tilde \Delta
< 1$ (the thin-shell limit is $\tilde \Delta \to 1$). The hollow
shell has an inner boundary at $R_{\rm{in}}=R \sqrt{2 \tilde
\Delta-1}$, while the filled geometries have no
inner boundary.
The condition $\tilde \Delta = 0.5$ demarcates the
sharp transition between the filled and hollow systems. In
finite-difference numerics, we section the interval $[0,R]$
($[R_{\rm{in}},R]$) for the filled (hollow) case into $2^d$
lattice sites and turn the relevant differential equation into a
generalized eigenproblem for a finite-size $(2^d + 1) \times (2^d
+1)$ matrix (see more details in Appendix
\ref{app:Finite_diference}). We choose a sufficiently large $d$
that guarantees the convergence of the solution. The data points
presented below are for $d=16$ (unless mentioned otherwise).

\begin{figure}[t]
\centering
  \includegraphics[width=6.5cm]{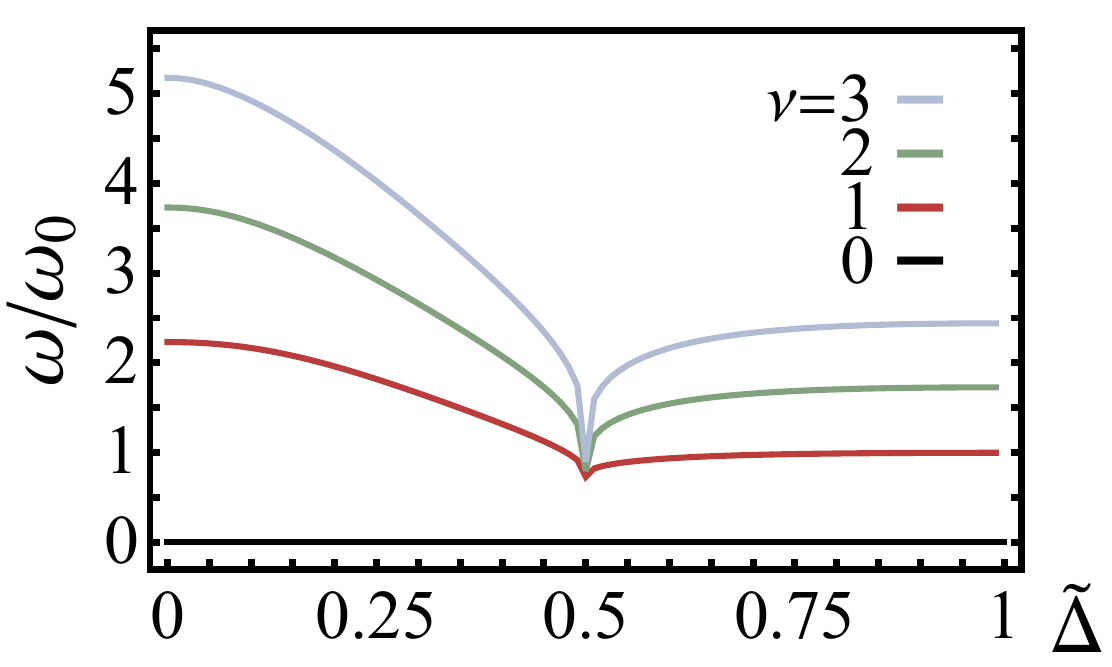}
\caption{(Color online) Oscillation frequencies $\omega$, found
via the hydrodynamic equation, Eq.~(\ref{eq:eigenproblem}), with
Thomas-Fermi equilibrium profiles, of the three lowest-lying
nonzero spherically symmetric ($l=0$) collective modes $\nu=1,2,3$
(dark red, green, and light blue curves, respectively) vs the
bubble-trap detuning $\tilde \Delta$ [given $\Delta/\Omega=1$ in
Eq.~(\ref{eq:bubble})]. The zero mode (black) is also presented
for comparison. As $\tilde \Delta$ increases, the BEC evolves from
a filled sphere $\tilde \Delta=0$ toward a hollow thin shell
$\tilde \Delta \to 1$, through a hollowing transition at $\tilde
\Delta=0.5$. In the sphere and thin-shell limits, the frequencies
agree with the exact solutions of Eqs.~(\ref{eq:spheremodes}) and
(\ref{eq:thin_shell_modes}), respectively. Around the transition
point, the collective-mode evolution is characterized by a dip in
frequency, which is singular in the Thomas-Fermi approximation due
to the appearance of a sharp new boundary.}
   \label{fig:3}
\end{figure}

\begin{figure}[t]
\centering
  \includegraphics[width=8.6cm]{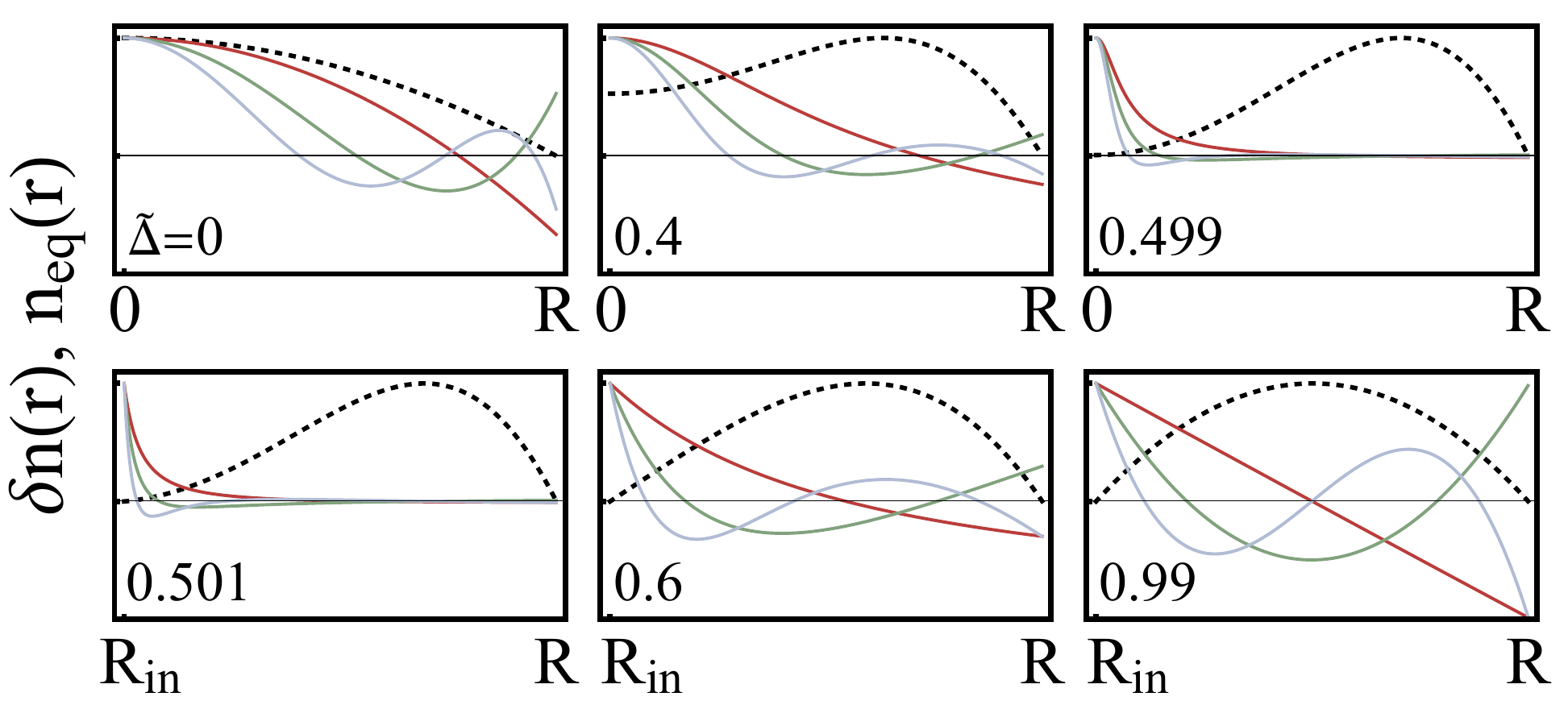}
\caption{(Color online) Normalized density deviation profiles
$\delta n(r)$ of the three physical breathing modes in
Fig.~\ref{fig:3} (same convention) and equilibrium profile
$n_{\rm{eq}}(r)$ (dashed curves). Individual panels correspond to
three filled cases $\tilde \Delta=0$, $0.4$, and $0.499$, and
three hollow cases $0.501$, $0.6$, and $0.99$ (as labeled inside).
Around the hollowing-out transition, the density deviation
profiles tend to concentrate on the spherical center. This is
energetically favorable and related to the development of
frequency dip, as discussed in Sec.~\ref{sec:VIB}.}
   \label{fig:4}
\end{figure}

In Fig.~\ref{fig:3}, we plot the oscillation frequencies for the
lowest-lying spherically symmetric ($\ell=0$) collective modes,
including a zero mode and three nonzero modes, as a function of
$\tilde \Delta$, representing the deformation
 of the condensate from a filled sphere ($\tilde
\Delta=0$) to a thin shell ($\tilde \Delta \to 1$). Note that the
zero mode $\nu=0$, corresponding to a constant density-deviation
profile, is not physically detectable for any $\tilde \Delta$. We
find that the curves do not cross each other and the set of
frequencies can distinguish between different stages of the
deformation. Frequency values in the two limiting cases are
consistent with those predicted for the filled-sphere and
thin-shell limits in Eqs.~(\ref{eq:spheremodes}) and
(\ref{eq:thin_shell_modes}), respectively. There is a frequency
dip in each of the three physical modes when the shell develops an
inner boundary at $\tilde \Delta = 0.5$. The frequencies
monotonically decrease (increase) with $\tilde \Delta$ if $\tilde
\Delta < 0.5$ ($>0.5$). We have confirmed the stability of this
dip structure as the continuum limit is approached (up to $d=23$).
The frequency dip presents a clear signature for the hollowing
transition in the system. The physical behavior giving rise to
this feature is explored below and in more detail in
Sec.~\ref{sec:VI}.

We now turn to the behavior of the density deviations $\delta n
(r)$ [recalling that the collective modes of the condensate
correspond to $\delta n({\bf{r}})=D(r)Y_{\ell m}(\theta,
\varphi)$] and their evolution as the system transitions from
filled sphere to thin shell. In Fig.~\ref{fig:4}, we show $\delta
n$ for the lowest three modes and the equilibrium density
$n_{\rm{eq}}$ at six values of $\tilde \Delta$. We see that at any
stage in the evolution between the sphere and the thin shell, the
number of nodes in $\delta n$ is always equal to the mode index
$\nu$, and the maximum amplitude occurs at the center (the inner
boundary $r=R_{\rm{in}}$) if the center is filled (hollow). When
$\tilde \Delta$ increases from zero, the central equilibrium
density starts to drop. As we approach the hollowing transition at
$\tilde \Delta =0.5$, the central density drops to zero and the
density deviations attain large amplitudes, localizing at the
center of the system. [Note that we still assume a strong
interaction such that the linearization of Eq.~(\ref{eq:delta_n})
is valid, i.e., for given scattering length $a_s$ and
characteristic interparticle spacing $a_d$, $a_d\sqrt{a_d/a_s}$ is
small compared with the length scale of the concentration.] For
$\tilde \Delta> 0.5$, i.e., the hollow-shell regime, the density
remains zero at $R_{\rm{in}}$ ($\neq 0$) and the density
deviations delocalize from $R_{\rm{in}}$ as $\tilde \Delta$
increases. We recover the Legendre polynomials in the
density-deviation profiles in the thin shell limit of $\tilde
\Delta=0.99$, as we have shown in Sec.~\ref{sec:IVB}.

We emphasize the frequency dip at the hollowing transition $\tilde
\Delta = 0.5$ and the concentration of density deviations around
the center as the main results of this section. In assessing the
generality of this frequency drop as a signature of a hollow
condensate, realistic factors such as moderate interaction
strength and nonsharp boundaries need to be taken into account.

\begin{figure}[t]
\centering
  \includegraphics[width=6.5cm]{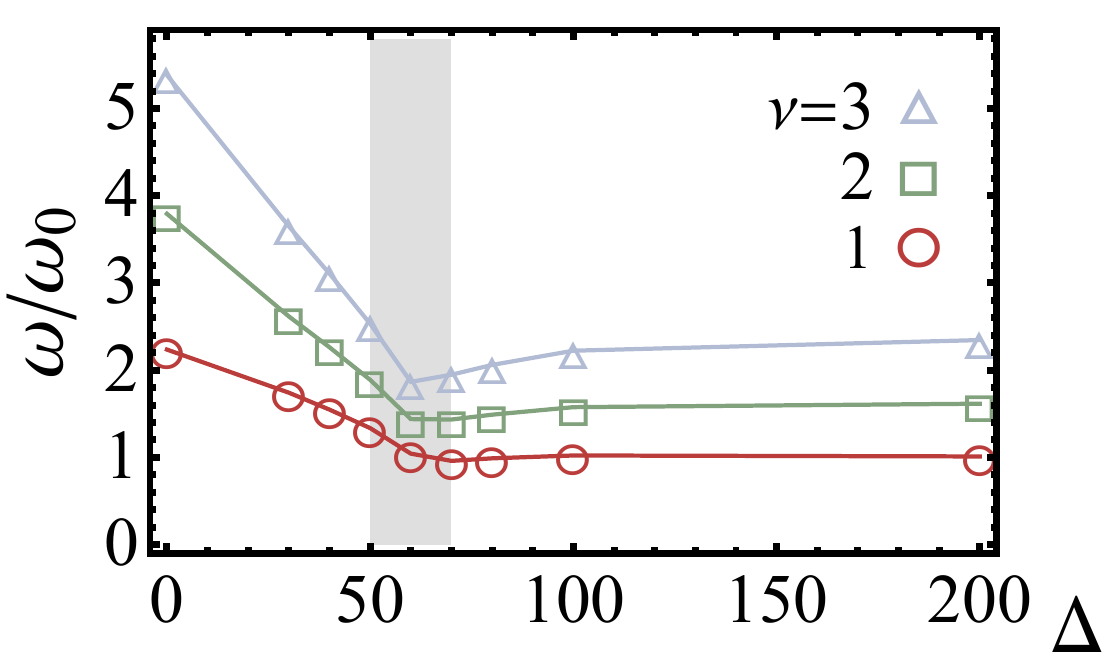}
\caption{(Color online) Oscillation frequencies $\omega$, found
via the hydrodynamic equation, Eq.~(\ref{eq:eigenproblem_noTF}),
with equilibrium profiles given by the numerical solution to the
GP equation, of the three lowest-lying nonzero spherically
symmetric ($l=0$) collective modes $\nu=1,2,3$ (circles, squares,
and triangles, respectively) vs the bubble-trap detuning $\Delta$.
The bubble-trap parameters are set by $\Delta/\Omega=1$ and the
interaction strength is $u=10^4$. The equilibrium profiles
corresponding to five of the data points are shown in
Fig.~\ref{fig:2}. Compared to the Thomas-Fermi results in
Fig.~\ref{fig:3}, the dip in frequency is softened but clearly
presents a hollowing transition region (shaded region) between
filled and hollow topologies.}
        \label{fig:5}
\end{figure}

As a first step in moving beyond the Thomas-Fermi approximation,
we recall that the hydrodynamic formalism for the BEC's dynamical
behavior can be applied to any equilibrium density profile, in
particular to the numerical ground state of the GP equation. In
Fig.~\ref{fig:5} we plot the frequencies of the $\nu=1,2,3$
spherically symmetric ($\ell=0$) collective modes calculated by
using hydrodynamic equation, Eq.~(\ref{eq:eigenproblem_noTF}),
with the GP equilibrium density profile found by employing the
imaginary-time algorithm. We see that the limiting behavior of the
collective mode frequencies is consistent with our discussion in
Sec.~\ref{sec:IV} regardless of whether we use the Thomas-Fermi
density profile or the more realistic numerical GP result. The dip
feature is still present, indicating the development of an inner
boundary, but the GP profiles give a less sharp frequency dip
compared to the Thomas-Fermi results. In fact, the transition
itself spreads across a region, as opposed to a single point. The
softened dip feature still reflects the transition region (shaded)
between filled and hollow behaviors. The sharpness of the
frequency dip is associated with the use of the Thomas-Fermi
approximation in which the density profile decreases to zero at
condensate edges in an abrupt fashion (as shown in
Fig.~\ref{fig:2}). We will discuss the relevant physics in details
in Sec.~\ref{sec:VIB}

\begin{figure}[t]
\centering
  \includegraphics[width=6.5cm]{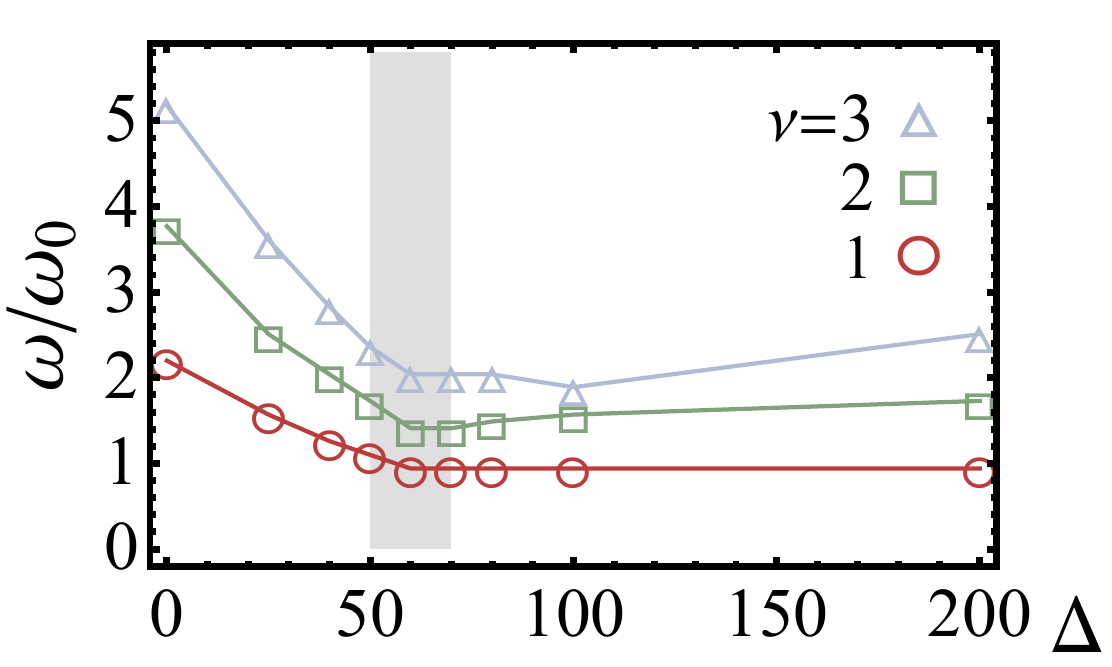}
\caption{(Color online) Oscillation frequencies $\omega$, found
via quench numerics, of the three lowest-lying nonzero spherically
symmetric ($l=0$) collective modes $\nu=1,2,3$ (same convention as
Fig.~\ref{fig:5}) vs the bubble-trap detuning $\Delta$. The
bubble-trap parameters are set by $\Delta/\Omega=1$ and the
interaction strength is $u=10^4$. In going beyond the Thomas-Fermi
approximation of Fig.~\ref{fig:5}, the singularity in the dip
feature around the hollowing transition is softened and spread
across the shaded region. However, the drop in frequency from the
filled  $\Delta=0$ point and the asymptote to the thin-shell limit
for large $\Delta$ persists and the spectrum demonstrates that the
collective-mode features through the topological transition are
robust beyond the hydrodynamic approximation.} \label{fig:6}
\end{figure}

To capture the most general physics beyond both the Thomas-Fermi
approximation and the hydrodynamic approach, we perform sudden
quench numerical simulations using the method described in
Sec.~\ref{sec:IIIB}. We simulate BECs throughout their evolution
in a bubble-trap from a filled sphere at $\Delta =0$ to a thin
shell at large values of $\Delta$. Figure \ref{fig:6} shows the
measured frequencies of the three lowest lying nonzero spherically
symmetric collective modes as a function of $\Delta$ ranging from
$0$ to $200$, holding $\Delta/\Omega=1$ and the total number of
particles $N$ fixed.

\begin{figure}[t]
\centering
  \includegraphics[width=8.6cm]{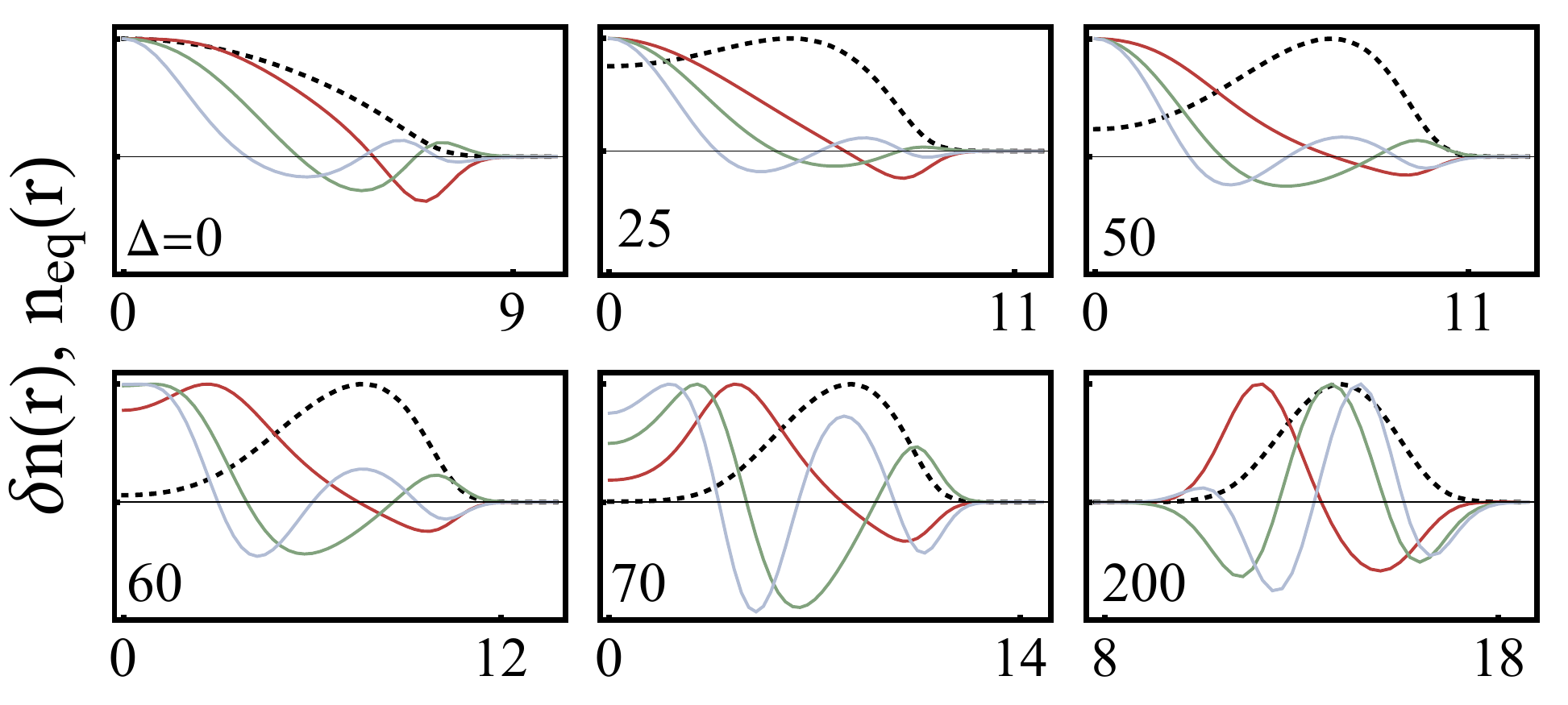}
\caption{(Color online) (Color online) Density deviation profiles
$\delta n(r)$ of the three breathing modes in Fig.~\ref{fig:6}
(same convention) and equilibrium profile $n_{\rm{eq}}(r)$ (dashed
curves). Individual panels correspond to $\Delta=0$, $25$, $50$,
$60$, $70$, and $200$ (as labeled inside). The radial coordinate
($x$ axis) is presented in the same units as Fig.~\ref{fig:2}.
Similar to the hydrodynamic results in Fig.~\ref{fig:4}, the
density deviation profiles tend to concentrate (but less
significantly so) around the spherical center in the hollowing
transition region. }
        \label{fig:7}
\end{figure}

We note the general agreement of the frequency spectrum found
using this method with the hydrodynamic results shown in
Figs.~\ref{fig:3} and \ref{fig:5}. This method also gives a more
rounded dip feature than that obtained using the Thomas-Fermi
approximation. This further supports the idea that the sharpness
of the frequency dip is related to the suddenness with which the
sphere hollows as a function of $\Delta$. We observe that the
frequency of the $\nu=1$ mode (shown in red) shows behavior very
similar to the results of the hydrodynamic approach with the
numerical ground state. As also found with the hydrodynamic
approach applied to the numerical ground state, the frequencies of
the three modes are well separated when the sphere hollows out,
i.e., at the value of $\Delta$ where the frequency dips.

Figure \ref{fig:7} shows the deviations of condensate density from
equilibrium in the numerical time evolution after a quench for
values of $\Delta$ that results in the deformation from the filled
sphere to the thin shell. We note these density deviations generally
reflect those found from the hydrodynamic approach in the
Thomas-Fermi approximation (Fig.~\ref{fig:4}), in that they also
show localization of the oscillations to the inner boundary for
$\Delta \approx 60$ where the frequency dip occurs as the center
of the system becomes hollow.

Arguably these numerical results are more representative of true
physical behavior of an experimental system than those obtained
using the hydrodynamic approach, which involves a number of
previously noted assumptions. The similarity between the two
results, especially prominent for hollow shells, justifies the use
of the more numerically efficient Thomas-Fermi hydrodynamic
approximation. Most importantly, it corroborates our prediction
that the dip in the frequency at the values of $\Delta$ where the
system transitions between the filled sphere and hollow shell is a
physical feature that could be observed in an experimental
setting.

\subsection{Evolution of modes having $\ell \neq 0$ in the bubble trap} \label{sec:VB}

We now examine how collective modes having $\ell \neq 0$ evolve as
the condensate is hollowed. We first note that in the eigenvalue
problem of Eq.~(\ref{eq:eigenproblem_noTF}) or
(\ref{eq:eigenproblem}), the nonzero $\ell$ brings in the term
$[V(R)-V(r)] \ell (\ell+1)/r^2 \propto n_{\rm{eq}}(r) \ell
(\ell+1)/r^2$, often called a ``centrifugal" term because of its
relationship with the angular momentum of the system. Without this
term ($\ell=0$), the right-hand side (RHS) of
Eq.~(\ref{eq:eigenproblem}) has only $r$-derivative terms, which
naturally guarantee a (unphysical) zero-frequency mode $\nu=0$
having a uniform density-deviation profile. We will see below that
this mode shifts to a finite frequency and hence becomes a
physical solution as the centrifugal term contributes for any
nonzero $\ell$. In addition, if $\ell$ is large enough such that
the centrifugal term dominates over the derivative terms, we shall
expect that the low-frequency modes are strongly affected by the
potential minima, which correspond to the boundary of the
condensate (where $n_{\rm{eq}}=0$). In this case, a filled
condensate with only an outer boundary is quite different from a
hollow condensate with both inner and outer boundaries, so the
high-$\ell$ collective modes can drastically change when the
bubble-trap system starts to hollow out.

\begin{figure}[t]
\centering
  \includegraphics[width=8.6cm]{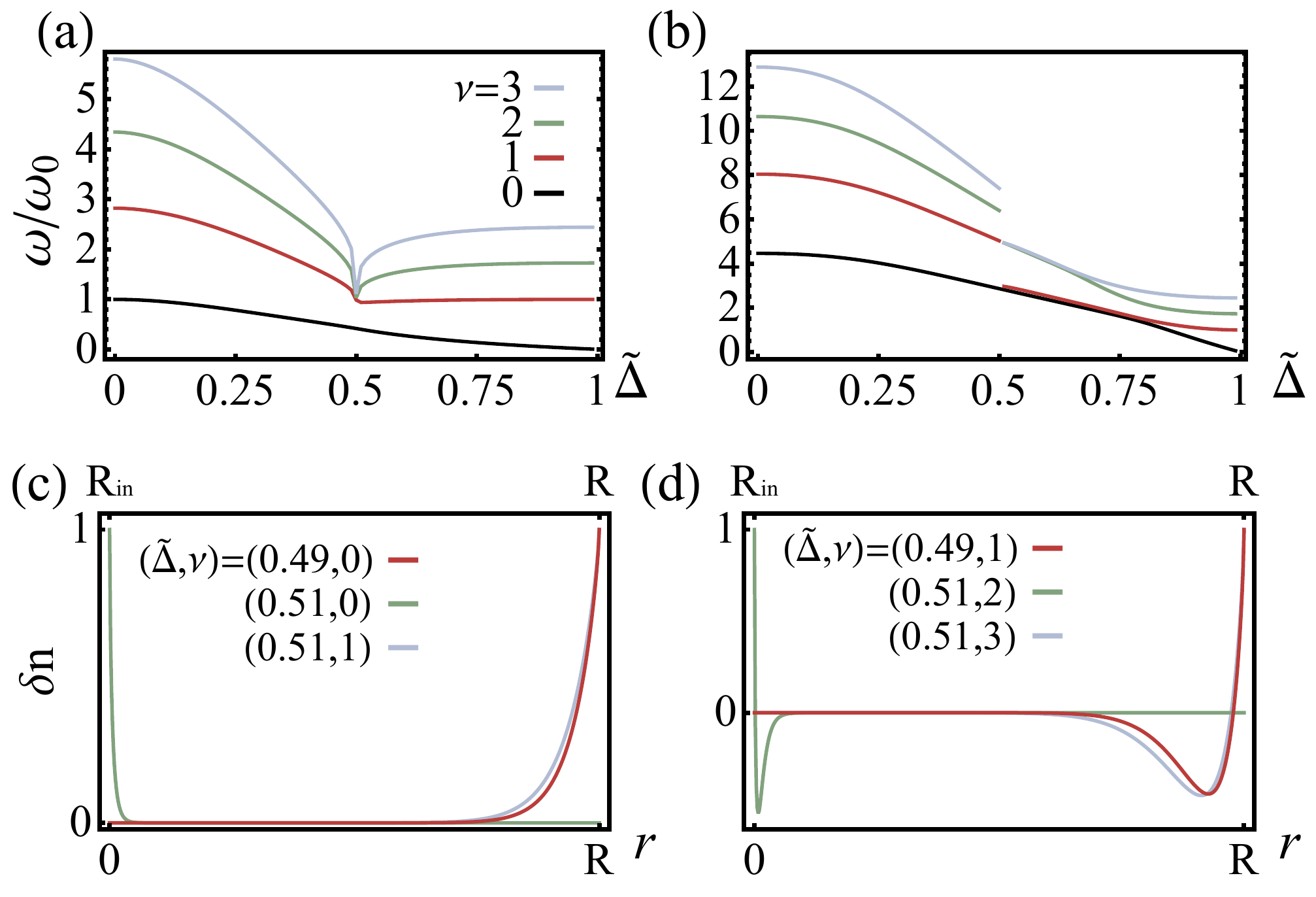}
\caption{(Color online) (a) [(b)] Oscillation frequencies $\omega$
vs $\tilde \Delta$ for $\ell=1$ ($\ell=20$) and $\nu=0,1,2,3$
(same convention as Fig.~\ref{fig:3}). The $\ell=1$ (low-$\ell$)
modes exhibit similar features, including the dip across the
hollowing transition $\tilde \Delta = 0.5$, as the $\ell=0$ modes
in Fig.~\ref{fig:3}, except that the $\nu=0$ mode here has nonzero
frequency. The $\ell=20$ (high-$\ell$) modes exhibit sudden drops
at the transition. (c) Density deviation profiles $\delta n$
corresponding to one mode $(\tilde \Delta,\nu) = (0.49,0)$ (dark
red) right before the hollowing (axis on bottom), and two modes
$(0.51,0)$ (green) and $(0.51,1)$ (light blue) right after it
(axis on top) in panel (b). (d) Same convention as panel (c)
except for modes $(0.49,1)$ right before the hollowing and
$(0.51,2)$ and $(0.51,3)$ right after it. Panels (c) and (d)
confirm that the high-$\ell$ modes are surface modes---with
density deviation localizing around the condensate surfaces.
Before the hollowing ($\tilde \Delta = 0.49$), the density
deviations of all modes localize around the only surface $r=R$,
while after the hollowing ($\tilde \Delta = 0.51$), half of them
remain around the outer surface $r=R$ and the other half
redistribute around the emerging inner surface $r =
R_{\rm{in}}\approx 0$. }
        \label{fig:8}
\end{figure}

We first study the $\ell =1$ case. Figure \ref{fig:8}(a) shows the
$\nu \le 3$ modes obtained from the hydrodynamic approach using
Thomas-Fermi equilibrium profiles. We see that the $\nu=0$ mode
now has a finite (and hence physically detectable) frequency that
is of the order ${O}(\omega_{\rm{sh}})$ on the filled-sphere side
and almost zero in the thin-shell limit, in agreement with the
limiting cases studied in Secs.~\ref{sec:IVA} and \ref{sec:IVB}.
The monotonically decreasing curve also shows how these two
different energy scales continuously connect through the
deformation. The three $\nu > 0$ modes exhibit qualitative
features and dip structures similar to those of the $\ell=0$ case,
except the filled-sphere frequencies in the $\ell=1$ case increase
by ${O}(\omega_{\rm{sh}})$ due to the angular oscillation.

We next study a large-$\ell$ case where the centrifugal term
dominates. Figure \ref{fig:8}(b) shows the same $\nu \leq 3$ modes
for $\ell=20$. We see that the $\nu=0$ curve still continuously
decreases to zero as we proceed from the filled sphere to the thin
shell. For the $\nu>0$ modes, the dip structure disappears, and
the curves exhibit a sudden drop upon the hollowing transition
$\tilde \Delta = 0.5$. The $\nu=1$ mode makes a drop to become
nearly degenerate with the $\nu=0$ mode, and the $\nu=2$ and
$\nu=3$ modes become equal and continue the evolution of the
$\nu=1$ mode before the drop. In fact, the $\nu$th mode right
before the hollowing point and the $(2\nu)$th and $(2\nu+1)$th
modes right after it are nearly degenerate such that the
$\nu$th-mode frequency curve appears to split into two upon the
hollowing transition.

We further investigate this splitting in the high-$\ell$ spectrum
by comparing the radial density-deviation profiles $\delta n(r)$
of three nearly degenerate modes---one just before the hollowing
($\tilde \Delta = 0.49$) and two just after the hollowing ($\tilde
\Delta = 0.51$). In Fig.~\ref{fig:8}(c) [Fig.~\ref{fig:8}(d)], we
plot the density deviation profiles for six specific modes in
Fig.~\ref{fig:8}(b). We see that before the hollowing transition,
the modes concentrate near the outer boundary. After hollowing one
of the nearly degenerate pair modes localizes near the outer
boundary while the other does near the newly created inner
boundary. This indicates that the inner minimum of the centrifugal
term provides a hollow condensate with additional degrees of
freedom for accommodating its collective modes. As the shell
becomes thinner, the pair of nearly degenerate modes further split
due to a coupling between them through the $r$-derivative terms in
Eq.~(\ref{eq:eigenproblem}).

Our findings show (i) how the frequencies of pure angular modes
($\nu=0$, $\ell \neq 0$) of a spherical condensate decrease during
the evolution to a shell condensate and (ii) how the high angular
modes qualitatively change when the inner boundary is created.
These results may find applications in nondestructive measurements
for determining the interior structre of a condensate. We will
further investigate the relation between the high angular modes
and the surface modes in Sec.~\ref{sec:VIC}.

\section{Physics of the hollowing-out signatures}\label{sec:VI}

In this section, we use our generalized radially shifted trapping
potential in order to study the signatures of the hollowing
transition over a family of spherically symmetric geometries,
These geometries have differing rate of decrease of the condensate
equilibrium density near the hollowing center. We provide physical
interpretations for the universal signatures of the hollowing
transition as well as for the distinct features that arise from
specific hollowing-out conditions. (i) We conclude that the
sharpness of the nonmonotonic spectral features, namely, the dip
at the hollowing transition depends on the rate of central
condensate density decay. As a result, the experimentally observed
spectrum should still exhibit a universal dip upon the transition.
However, the dip is less sharp than the predictions using
Thomas-Fermi approximation due to the more realistic equilibrium
density distribution with continuous ``tails'' at the boundaries.
(ii) Additionally, while the sudden drop of the surface-mode
spectrum is a universal signature due to the emergence of any
additional surface upon the hollowing transition, the double
degeneracy between inner and outer surface modes of a hollow shell
only occurs for specific geometries, including that of the
bubble-trap confinement.

In Sec.~\ref{sec:VIA}, we discuss the general radially shifted
potential given in Eq.~(\ref{eq:gt_potential}) having two
parameters: $\gamma$, which tunes the BEC geometry from filled
sphere to hollow shell, and $\alpha$, which controls the decay of
the equilibrium density near the center of the system as it
hollows. In Sec.~\ref{sec:VIB}, by varying the behavior of the
central density, we show that the radial collective mode spectrum
exhibits a universal dip at the hollowing transition, but that the
sharpness of the dip depends on $\alpha$. We also use an energetic
argument followed by a variational calculation to further
understand the frequency dip and the associated concentration of
density deviations near the hollowing center. In
Sec.~\ref{sec:VIC}, we show that while the hollowing transition
always leads to additional surface modes on the inner surface, the
mode frequencies depend on the surface stiffness, which in turn
depends on $\alpha$. The frequency degeneracy of the inner and
outer surface modes, which we have observed in the bubble trap,
corresponds to a special case where $\alpha = 2$.  In
Sec.~\ref{sec:VID}, we study the hollowing-out physics in a 2D
system and show that the collective-mode features are dimension
independent.

\subsection{General radially shifted potential}\label{sec:VIA}

Here we consider the potential given in
Eq.~(\ref{eq:gt_potential}):
\begin{eqnarray*}
V_{\rm{gt}}(r) = \frac{1}{2}m\omega^2_{\rm{gt}} R^2 S_l^2 {\left[
{{{\left( {\frac{r}{R}} \right)}^\alpha } - \gamma } \right]^2},
\end{eqnarray*}
where $\gamma$ is a dimensionless parameter that tunes the
condensate between sphere and shell geometries, $\alpha$ tunes the
condensate equilibrium density profile near the hollowing center,
and $R$ specifies a characteristic size.

To understand the geometry of a condensate subject to
this trap, we focus on the Thomas-Fermi density profile,
\begin{eqnarray}
n_{{\rm{eq}}}^{{\rm{gt}}}({\bf{r}}) = n_0^{{\rm{gt}}}\left[ {1 -
{{\left( {\frac{r}{R}} \right)}^\alpha }} \right]\left[ {{{\left(
{\frac{r}{R}} \right)}^\alpha } - (2\gamma  - 1)}
\right],\label{eq:gt_TF_profile}
\end{eqnarray}
with an overall magnitude $n_0^{{\rm{gt}}} = {m{(\omega
_{\rm{t}}^{{\rm{gt}}})^2}{S_l^2 R^2}}/{(2U)}$. The condensate has
its outer boundary at $r=R$ (where $n_{{\rm{eq}}}^{{\rm{gt}}}$
vanishes) and its density maximum at $r=\gamma ^{1/\alpha} R
\equiv r_0^{\rm{gt}}$. This density profile shows that the general
trap produces a filled sphere condensate for $\gamma=0$, a
thin-shell condensate for $\gamma \to 1$, and transitions between
them as $\gamma$ is varied.

\begin{figure}[t]
\centering
  \includegraphics[width=8.6cm]{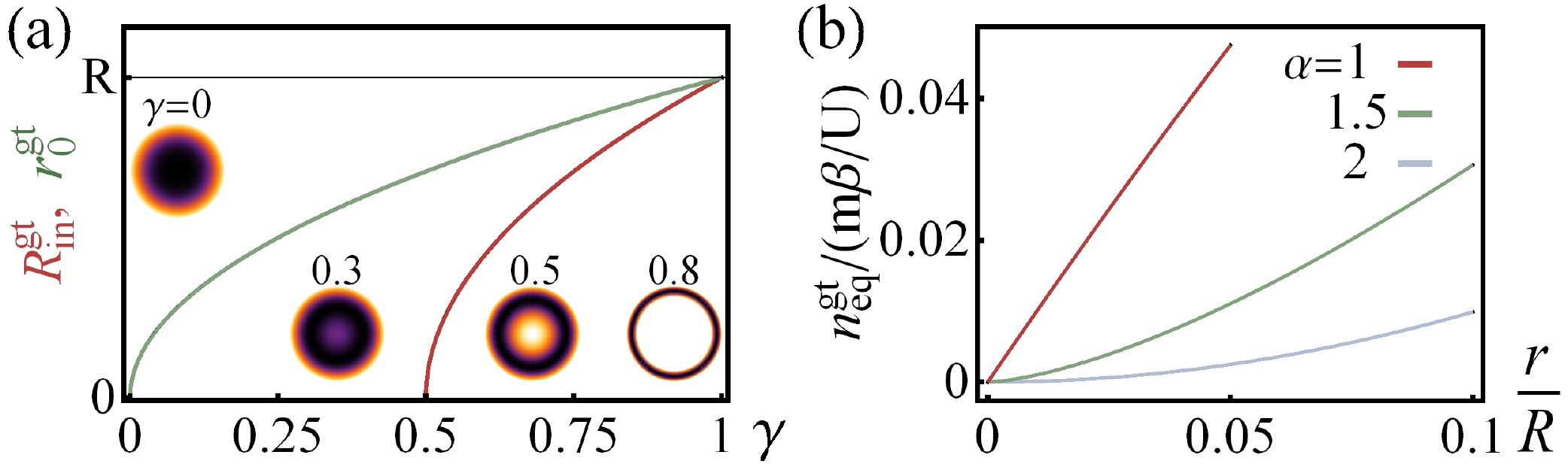}
\caption{(Color online) (a) Thomas-Fermi inner boundary
$R_{{\rm{in}}}^{{\rm{gt}}}$ (dark red) and maximum-density
position $r_0^{\rm{gt}}$ (green) vs $\gamma$ for a condensate in
the general trap, given by Eq.~(\ref{eq:gt_potential}) for $\alpha
= 2$. Insets from left to right: schematic density profiles of the
condensate at $\gamma=0$, $0.3$, $0.5$, and $0.8$, respectively.
(b) Thomas-Fermi density profiles $n_{{\rm{eq}}}^{{\rm{gt}}}(r)$
around the center for $\alpha=1$ (dark red), $1.5$ (green), and
$2$ (light blue) at the hollowing transition $\gamma=0.5$.}
        \label{fig:9}
\end{figure}

At $\gamma=0$, the condensate forms a fully filled sphere with
density $n_{{\rm{eq}}}^{{\rm{gt}}} \propto 1 - (r/R) ^{2 \alpha}$,
exhibiting monotonically decreasing density as one moves radially
out from the origin. As $\gamma$ starts to increase from zero, the
maximum density position shifts to a finite radius. For $0 <
\gamma < 0.5$, the condensate density at the origin,
$n_{{\rm{eq}}}^{{\rm{gt}}}(0) = n_0^{{\rm{gt}}}(1 - 2\gamma )$,
remains finite and a local minimum. At $\gamma = 0.5$, the
condensate density at the origin becomes zero, signaling the
transition from filled to hollow geometry. For $\gamma
> 0.5$, we have a hollow shell system with an inner boundary at
$R_{{\rm{in}}}^{{\rm{gt}}} = {(2\gamma -
1)^{1/\alpha }}R$. As $\gamma \to 1$, the condensate
approaches the thin-shell limit, where
${V^{{\rm{gt}}}}(\mathbf{r})$ can be approximated by the
thin-shell potential of Eq.~(\ref{eq:Vsh}) having parameter values
$r_0 = r_0^{\rm{gt}}$ and $\omega _{\rm{t}}^{{\rm{sh}}} = {\gamma
^{1 - 1/\alpha }} \alpha \omega _{\rm{t}}^{{\rm{gt}}} \sim \alpha
\omega _{\rm{t}}^{{\rm{gt}}}$ (and hence
$c=1/[\gamma^{-1/\alpha}-1] \gg 1$).

In Fig.~\ref{fig:9}(a), we plot $R_{{\rm{in}}}^{{\rm{gt}}}$ and
$r_0^{\rm{gt}}$ vs $\gamma$ and schematic density profiles
(insets) for $\alpha=2$ (quartic double-well) case. The curves
show the continuous evolution of the condensate peak density and
inner boundary, while the schematics show the density at four
stages of the evolution: sphere ($\gamma=0$), thick shell with
filled center ($0.3$), thick shell at the hollowing transition
($0.5$), and thin shell ($0.8$). The variable $\alpha$ determines
the power law of the equilibrium profile's growth at the center of
the system as $n_{{\rm{eq}}}^{{\rm{gt}}} \propto r ^{\alpha}$. In
Fig.~\ref{fig:9}(b), we plot the Thomas-Fermi profiles near the
center of the system at the hollowing transition ($\gamma=0.5$)
for various $\alpha$. For the bubble-trap potential, the
Thomas-Fermi density corresponds to $\alpha=2$, while the
numerical GP solution behaves as $\alpha<2$ since the condensate
wave function has a continuous ``tail" at its boundaries.

In Sec.~\ref{sec:V}, we presented results on the collective mode
spectrum for the breathing modes in the Thomas-Fermi approximation
and using the numerical ground state density, in the bubble-trap
potential. We found some universal features (e.g., a dip in the
frequency spectrum), as well as some notable differences (e.g.,
the sharpness of the dip). We see that analyzing the collective
modes in the general potential of Eq.~(\ref{eq:gt_potential}) and
varying $\alpha$ will allow us to distinguish universal signatures
of the hollowing transition from those that depend on the central
density profile. In the following two sections, we provide a
detailed analysis of the effect of $\alpha$ on the spectra of the
radial and surface collective modes. We note that for $\alpha \le
1$, the density profile has a kink (discontinuity in its
derivative) at the center, which leads to divergent kinetic energy
density $|\partial_r \psi(0)|^2$ and is hence unphysical.
Consequently, below we only consider $\alpha
> 1$.

\subsection{Spherically symmetric modes at the hollowing transition} \label{sec:VIB}

From both the hydrodynamic treatment and the sudden quench
numerics, we have observed a dip in the frequency spectrum of
radial collective modes when the spherical BEC in bubble trap
starts to hollow out at its center. However, the results based on
a Thomas-Fermi equilibrium density profile show a sharper dip (as
in Fig.~\ref{fig:3}) than those based on a profile from the GP
equation (as in Figs.~\ref{fig:5} and \ref{fig:6}). In addition,
we find that the former have density deviations more localizing
near the trap center ($r=0$) than the latter for values of
$\Delta$ near the hollowing transition. Because the most
significant difference between the Thomas-Fermi and GP equilibrium
density profiles lies in the central region at the hollowing
transition, and the concentrated density deviations are mainly
determined by the equilibrium density near the center [through
Eq.~(\ref{eq:delta_n_no_TF2})], one can hypothesize that the shape
of central equilibrium density and the sharpness of the frequency
dip are closely related. In this section, we study a convenient
model using the general-trap profile of
Eq.~(\ref{eq:gt_TF_profile}) to verify this relationship and
confirm that the GP equilibrium density profile leads to a less
sharp frequency dip, which should better characterize results in
real experimental systems. We also provide evidence based on
energetics points and supported by a variational calculation that
the frequency dip is always accompanied by the concentration of
density deviations at the hollowing transition.

The shape of the central density of a hollowing condensate can be
characterized by a power-law index $\alpha$, such that
$n_{\rm{eq}}(r) \propto r^\alpha$. As we have shown in
Sec.~\ref{sec:VIA}, the general-trap potential can both tune
through the sphere-to-shell evolution and control the index
$\alpha$, thus making it ideal for studying the effects of central
density during the hollowing transition. Tuning $\alpha$ at
$\gamma=0.5$, the transition point between filled and hollow
topologies, can help theoretically isolate the effects of the
central density growth ($\propto r^\alpha$). We comment that for a
bubble-trap potential, the central Thomas-Fermi profile varies as
$r^2$ when the system is hollowing out. However, the solution of
the GP equation is distinctly different due to relatively smooth
``tail" near the center associated with a relatively large kinetic
energy contribution in this region, which is ignored in the
Thomas-Fermi approximation. Therefore a more realistic description
of the frequency dip in a bubble traps can be captured by the
density growth of modified profile with $\alpha < 2$ than by the
original Thomas-Fermi one.

\begin{figure}[t]
\centering
  \includegraphics[width=8.6cm]{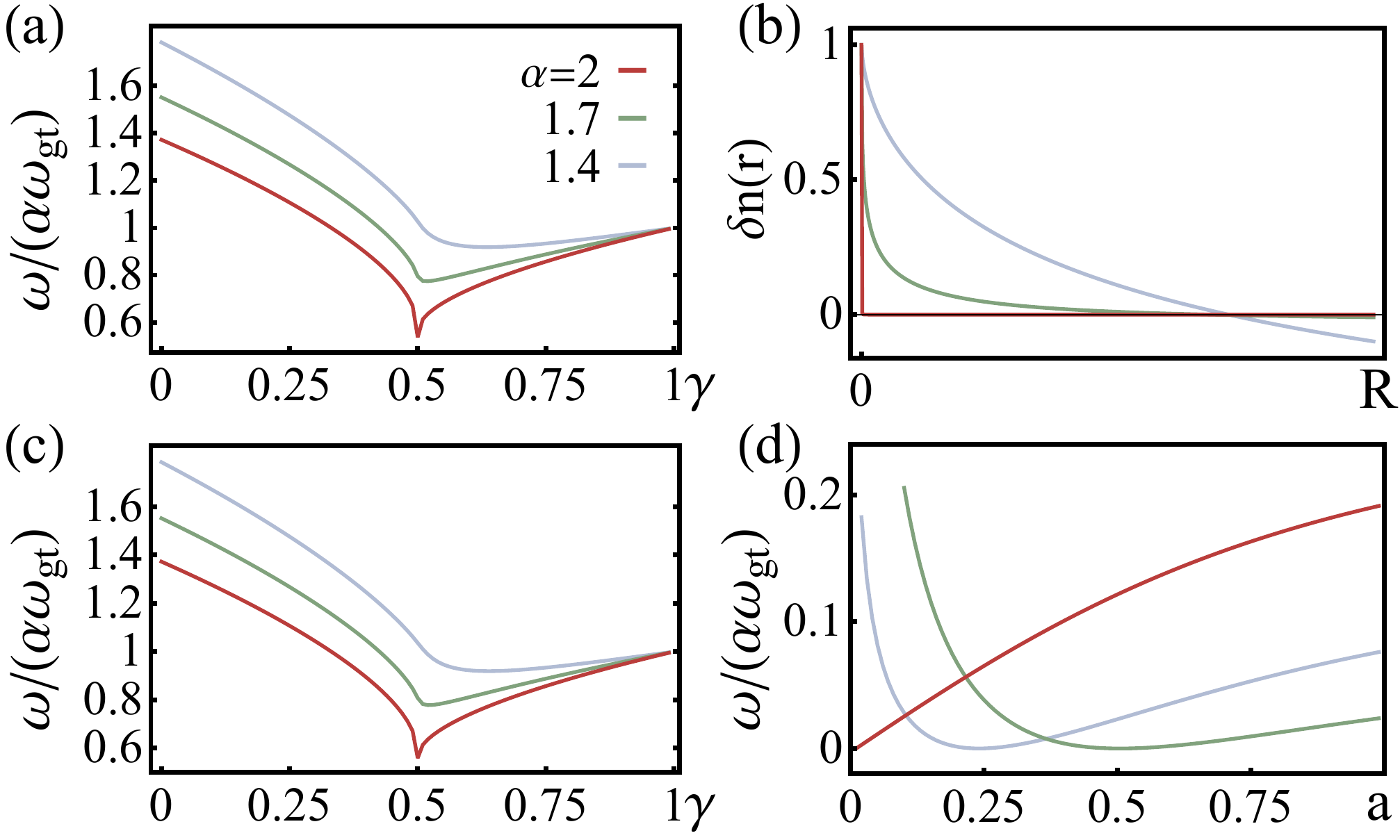}
\caption{(Color online) (a) Oscillation frequencies $\omega$ for
the lowest nonzero breathing mode ($\nu=1$, $\ell=0$) as a
function of $\gamma$ for $\alpha=2$ (dark red), $1.7$ (green), and
$1.4$ (light blue), obtained by solving the hydrodynamic equation
for the general-trap Thomas-Fermi profiles in
Eq.~(\ref{eq:gt_TF_profile}). (b) The corresponding density
deviation profiles $\delta n$ at the hollowing transition, $\gamma
= 0.5$. (c) Oscillation frequencies $\omega$ obtained from the
variational method. (d) Variational frequency function at
$\gamma=0.5$. Panels (b)--(d) have the same convention as panel
(a). Panels (a) and (b) show the dependence of the frequency dip
sharpness and the concentration of density deviation profiles on
the growth rate of the equilibrium density profile from the
hollowing-out center. Panels (c) and (d) show that the dependence
originates from the orthonormality and energy minimization of
collective mode. }
        \label{fig:10}
\end{figure}

Figure \ref{fig:10}(a) shows the frequency of the lowest nonzero
radial collective mode ($\nu=1$, $\ell=0$) as a function of
$\gamma$ at various $\alpha \le 2$, obtained from the hydrodynamic
approach. The frequency dip is clearly identifiable near
$\gamma=0.5$ in all three curves, but its sharpness changes with
the growth rate $\alpha$. More precisely, we see that for $\alpha
< 2$, the frequency-dip structure is smoother and the
minimum-frequency point shifts slightly to $\gamma > 0.5$. In
Fig.~\ref{fig:10}(b), we plot the density deviation profiles
$\delta n(r)$ corresponding to each $\alpha$ in
Fig.~\ref{fig:10}(a) at $\gamma = 0.5$. We see that the density
deviations become less concentrated at the center as $\alpha$
decreases from 2. As a result, the $\alpha=2$ density growth,
leading to sharp frequency dips and highly concentrated density
deviations, agrees with the results for a bubble-trap system with
Thomas-Fermi density profiles (as in Figs.~\ref{fig:4} and
\ref{fig:5}). On the other hand, the $\alpha<2$ density growth,
leading to less sharp frequency dip and less concentrated density
deviations, agrees with the results using the GP density profiles
(as in Figs.~\ref{fig:5}, \ref{fig:6}, and \ref{fig:7}).
Therefore, the general-trap model captures the main features of
the spherically symmetric collective mode frequency spectra during
the hollowing transition. We conclude that for a bubble-trap
system, using the Thomas-Fermi profile in a general trap and
setting $\alpha < 2$ can mimic the more realistic GP profile
without the need for numerics.

We next discuss the relationship between the frequency dip and the
concentration of density deviations at the hollowing transition.
To first provide an energetic argument for interpreting the
frequencies of the collective modes, the $\nu$th collective mode
has a radial density deviation profile that is orthogonal to all
the lower-frequency modes and minimizes the energy (frequency).
This is a direct result of the orthonormality of the
eigen-solutions of Eq.~(\ref{eq:delta_n_no_TF2}),
\begin{eqnarray*} -\frac{mS_l^2}{U}\omega^2\delta n=\nabla n_{ \rm{eq}}\cdot
\nabla\delta n+n_{ \rm{eq}}\nabla^2\delta n.
\end{eqnarray*}
We see from Eq.~(\ref{eq:delta_n_no_TF2}) that the equilibrium
density profile contributes to two terms, $n_{\rm{eq}}$ and
$\nabla n_{\rm{eq}}$. For most of the evolution from a filled
sphere to a thin shell, these two terms are not simultaneously
zero---we note that at the filled center $\nabla n_{\rm{eq}} (0) =
0 $ but $n_{\rm{eq}} (0) \neq 0$, and at the inner boundary of a
fully hollow shell, $n_{\rm{eq}} (R_{\rm{in}})= 0$ but $\nabla
n_{\rm{eq}} (R_{\rm{in}}) \neq 0 $. At least one of these two
terms increases the energy (frequency) and hence disfavors the
concentration of density deviations.

However, when the system is at the hollowing transition, both of
these terms can become very small in the central region, i.e., the
whole differential operator on RHS of
Eq.~(\ref{eq:delta_n_no_TF2}) nearly vanishes. In this case, a
concentration of density deviation profiles near the center of the
system will have a very small energy (frequency) expense and
should hence be favored in the low-frequency modes of the system
(any density deviation profile spreading away from the center will
instead result in a higher frequency). This very small
contribution from the RHS of Eq.~(\ref{eq:delta_n_no_TF2}) also
leads the mode frequencies to be much lower than those in the
filled sphere and the hollow shell (where either $n_{\rm{eq}}$ or
$\nabla n_{\rm{eq}}$ contributes)---hence forming a dip in the
frequency spectrum.

We further verify this energetic argument by solving
Eq.~(\ref{eq:eigenproblem_noTF}) for the first breathing mode
($\nu=1,\ell=0$) by a variational calculation. We consider a
variational density deviation profile $D_1(r)$ that has one node
($\nu=1$) and is orthogonal to the zero mode ($\nu=0$) with a
uniform density deviation $D_0$. We then solve for $D_1(r)$ by
minimizing the oscillation frequency functional. We adopt a
variational ansatz,
\begin{eqnarray}
{D_1}(r) = \frac{1}{{{{(r/a)}^2} + 1}} - b. \label{eq:var_D1}
\end{eqnarray}
This density deviation profile has a node at $r = a\sqrt {{b^{ -
1}} - 1}$. Orthogonality to the zero mode can be guaranteed by
choosing $b$ such that $\int\limits {{D_0}{D_1}(r){r^2} dr}=0$.
With this ansatz, we compute the frequency $\omega$ as the
expectation value of the differential operator of
Eq.~(\ref{eq:eigenproblem_noTF}), namely,
\begin{eqnarray}
\frac{mS_l^2}{U}{\omega ^2} = \frac{\int\limits n_{\rm{eq}}
{(\partial_r D_1)}^2{r^2}dr}{\int\limits {D_1^2{r^2}dr}}.
\label{eq:var_frequency}
\end{eqnarray}
The solution can then be obtained by minimizing $\omega$ with
respect to the only free parameter $a$.

In Fig.~\ref{fig:10}(c), we plot the variational results for
various values of $\alpha$. We see that the frequency dips upon
the hollowing out and that the dependence of dip sharpness on the
density profile are identical to those in Fig.~\ref{fig:10}(a)
given by the hydrodynamic equation. In Fig.~\ref{fig:10}(d), we
plot the frequency of this mode at the hollowing transition,
$\gamma=0.5$, as a function of the variational parameter $a$,
which is also the full width at half maximum of the density
deviation profile. We see that for the $\alpha=2$ case (with a
sharp dip in the spectrum), $a \to 0$ directly leads to the
minimization of frequency, and that therefore the density
deviation will be localized to the center. For the $\alpha<2$
cases (with a smooth dip), the frequency minimization occurs at
nonzero $a$, giving density deviations with some small, but
nonzero, width. These variational results not only agree with
those given the hydrodynamic equation but also corroborate the
energetic argument that the density deviations localizing near the
hollowing center is directly related to the dip in frequency.

We conclude that independent of the specific shape of the
confining trap, any condensate system that transitions between a
filled sphere and a hollow shell will display a universal
signature of its hollowing. Specifically, radial collective mode
spectrum exhibits a frequency minimum at the hollowing transition.
This is a rather remarkable result since it allows one to deduce
the appearance of a hollowing region deep within the condensate by
imaging  lowest-lying collective excitations of the system, which
can be observed on even the outer surface, possibly even
nondestructively.

\subsection{Surface modes at the hollowing transition}\label{sec:VIC}

We now turn to large-$\ell$ collective modes, which manifest
themselves as distortions localized near the boundary as the
condensate's surface modes~\cite{AlKhawaja1999}. The localization
of these modes is due to the dominant centrifugal term
$\ell(\ell+1)n_{\rm{eq}}$ in Eq.~(\ref{eq:eigenproblem_noTF}),
which becomes small only near the boundary, where $n_{\rm{eq}}=0$.
For our hollowing system, the significant feature is that a hollow
spherical shell supports similar minima in this term on both its
inner and outer surfaces. When the system becomes hollow, the
creation of the new inner boundary enables the localization of
large-$\ell$ density deviations to the inner surface. The
availability of the emerging surface doubles the surface mode
spectrum such that half of modes remain at the outer surface but
the other half redistribute to the inner surface.

The surface mode frequencies are determined by the properties of
the surfaces, as discussed in Sec.~\ref{sec:IVC} and seen
specifically in Eq.~(\ref{eq:dispersion}). Focussing on the
general-trap potential in the shell region, $\gamma \ge 0.5$, we
linearize the trapping potential and the corresponding
Thomas-Fermi equilibrium density close to the inner and outer
boundaries, $R_{\rm{in}}$ and $R_{\rm{out}}=R$, respectively, in
order to find
\begin{eqnarray}\label{eq:surfacemodeslin}
n_{{\rm{eq}}}^{{\rm{TF}}}({x_{{\rm{in,out}}}}) =  -
\frac{{{F_{{\rm{in,out}}}}}}{U}{x_{{\rm{in,out}}}},
\end{eqnarray}
with $F_{\rm{in,out}}=-\nabla V_{\rm{gt}}(R_{\rm{in,out}})$ and
$x_{\rm{in,out}} \le 0$ the local variable pointing along the
direction of $F_{\rm{in,out}}$. Employing the equilibrium density
profile of Eq.~(\ref{eq:gt_TF_profile}) to leading order in
$|r-R_{\rm{in,out}}|$ in the hydrodynamic equation of motion,
Eq.~(\ref{eq:hydrox}) yields the wave number associated with each
surface mode given by $q_{\rm{in,out}}=\ell/R_{\rm{in,out}}$ and
\begin{eqnarray}\label{eq:surfacemodesomega}
&&\omega^2_{\rm{out}}=\alpha \omega_{\rm{gt}}^2 S_l^2 \ell
(1-\gamma)(2\nu_{\rm{in, out}}+1), \nonumber\\
&&\omega^2_{\rm{in}}=\alpha \omega_{\rm{gt}}^2 S_l^2 \ell (2
\gamma-1)^{1-2/\alpha}(1-\gamma)(2\nu_{\rm{in, out}}+1).
\end{eqnarray}
As before, indices $\nu_{\rm{in}}$ and $\nu_{\rm{out}}$ in these
expressions count the nodes of radial oscillations confined to the
inner or the outer boundary surface, respectively (rather than
counting the total number of radial nodes $\nu_{\rm{in}} +
\nu_{\rm{out}}$ across the entire shell which is denoted by
$\nu$).

\begin{figure}[t]
\centering
  \includegraphics[width=8.6cm]{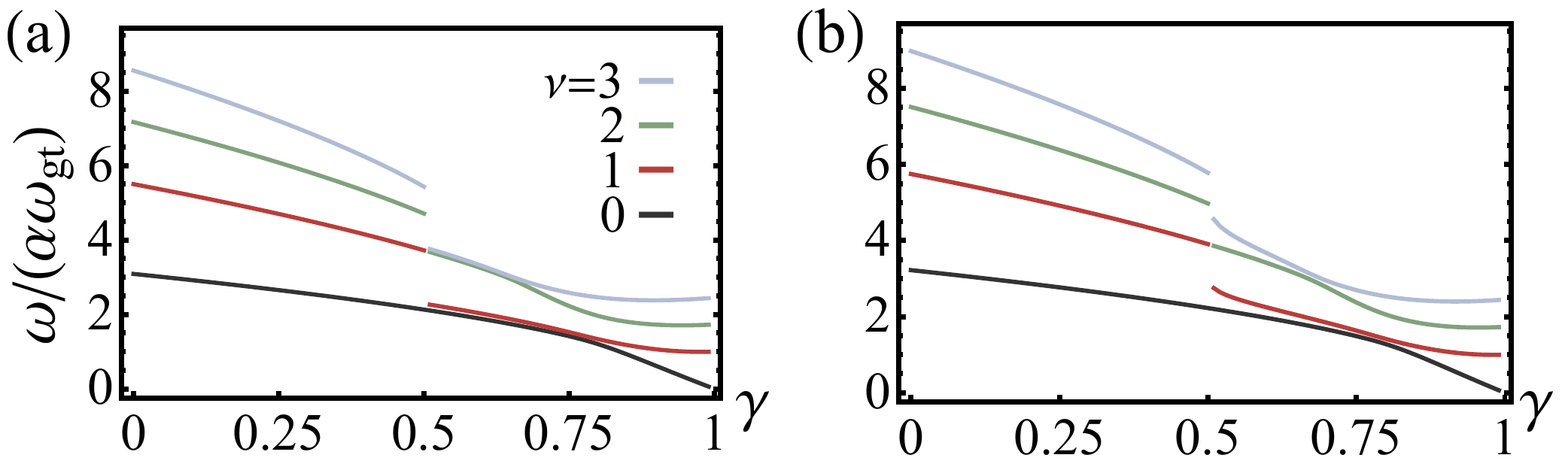}
\caption{(Color online) (a) [(b)] Oscillation frequencies $\omega$
of high-angular-momentum modes $\ell=20$ in the general-trap case
vs $\gamma$ for $\nu=0,1,2,3$ (same convention as
Fig.~\ref{fig:3}) and $\alpha=2$ ($1.85$). Comparing (a) and (b),
we see that the sudden drop of surface-mode frequency is a
universal feature for the hollowing transition, but the degeneracy
between outer and inner surface modes of hollow condensates only
occurs at $\alpha=2$. }
        \label{fig:11}
\end{figure}

We note that for the $\alpha=2$ case [including the bubble-trap
one in Fig.~\ref{fig:8}(b)], there is a degeneracy in the
frequency of surface modes at the inner and outer surfaces:
$\omega^2_{\rm{in}}=\omega^2_{\rm{out}}$. If $\alpha \neq 2$, this
double degeneracy no longer exists, but the frequency spectrum
still exhibits a clear drop due to the mode redistribution to the
newly emerging inner surface. In Fig.~\ref{fig:11}(a)
[Fig.~\ref{fig:11}(b)], we plot the oscillation frequency spectrum
of the high-angular-momentum mode $\ell=20$ for the general-trap
case with $\alpha=2$ (1.85). We see that near the hollowing-out
region $\gamma \gtrsim 0.5$, the double degeneracy occurs at
$\alpha=2$ but no longer exists at $\alpha = 1.85$, in agreement
with the surface-mode characteristics in
Eq.~(\ref{eq:surfacemodesomega}). In the thin-shell region $\gamma
\lesssim 1$, the modes strongly couple with each other and are no
longer degenerate (neither localize near the surface anymore). The
restructuring of the large-$\ell$ surface mode spectrum is a
direct and universal signature of a newly emerging surface and
hence the hollowing transition.

\subsection{The hollowing transition in two dimensions}\label{sec:VID}
Here, we investigate whether the hollowing-out physics we have
found for a 3D spherical system also occurs in an analogous 2D
geometry. We obtain the collective-mode frequency spectra of a
condensate in a 2D bubble trap, which realizes the hollowing
transition from a filled disk to a hollow ring. The trap takes the
same form as the 3D bubble trap of Eq.~(\ref{eq:bubble}) except
the coordinates are restricted to the $x$-$y$ plane ($z=0$). The
2D density deviation profile for a circularly symmetric condensate
has the form $\delta n({\bf{r}})=D(r)e^{i \ell \phi}$. Employing
this form in hydrodynamical equation of
Eq.~(\ref{eq:delta_n_no_TF2}) yields the differential
eigenequation,
\begin{eqnarray}\label{eq:eigenproblem_noTF_2D}
\frac{mS_l^2}{U}{\omega ^2 r}D = - \frac{d}{dr} \left( r
n_{\rm{eq}} \frac{d D}{dr} \right )+\frac{\ell^2 n_{\rm{eq}}}{r}D.
\end{eqnarray}

\begin{figure}[t]
\centering
  \includegraphics[width=8.6cm]{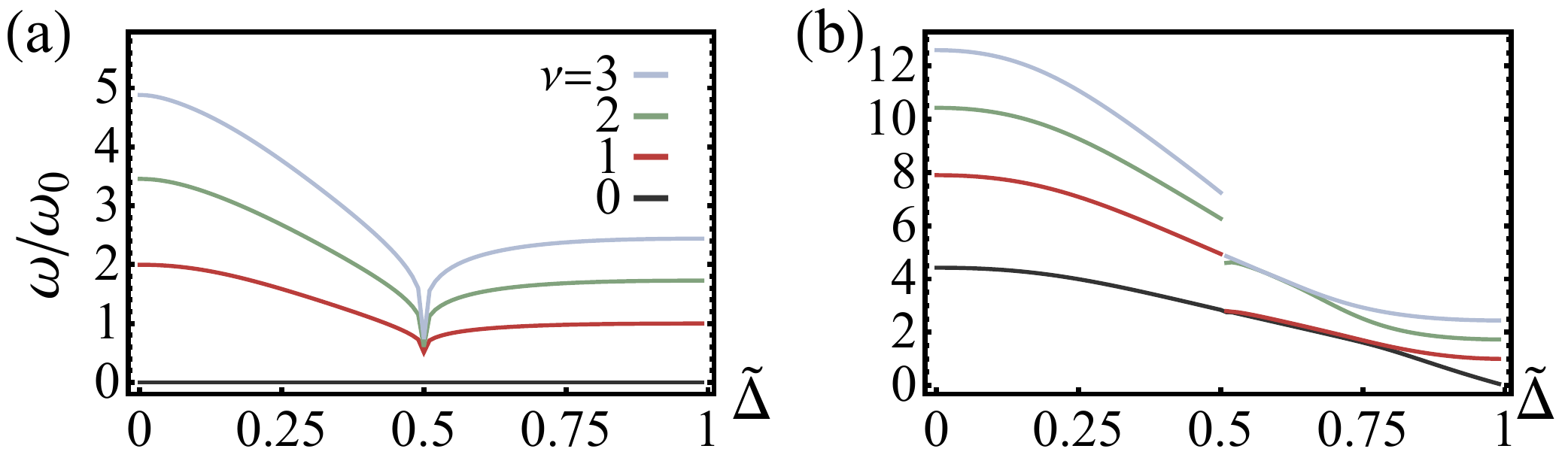}
\caption{(Color online)(a) [(b)] Oscillation frequencies $\omega$
of a 2D condensate vs 2D bubble-trap parameter $\tilde \Delta$ for
breathing modes $\ell=0$ (edge modes $\ell=20$) and $\nu=0,1,2,3$
(same convention as Fig.~\ref{fig:3}). Both panels show the same
collective-mode features in the filled-to-hollow evolution and
upon the hollowing transition as in 3D cases, indicating that the
hollowing-out physics is dimension independent.}
        \label{fig:12}
\end{figure}

In Fig.~\ref{fig:12}, we show the 2D collective-mode spectrum for
the Thomas-Fermi equilibrium density profile $n_{\rm{eq}}$, which
exhibits a sharp hollowing transition at $\tilde \Delta = 0.5$
(just as in the 3D case). Panel (a) shows that the breathing mode
($\ell=0$) frequencies are nonmonotonic with $\tilde \Delta$ and
develop a sharp dip at the 2D hollowing transition. A variational
analysis confirms that the frequency dip in this 2D system can be
derived from the orthonormality and energy minimization of the
collective modes and is associated with the density deviation
concentrating in the central hollowing region. Figure
\ref{fig:12}(b) shows that the high-$\ell$ ($\ell=20$) mode
frequencies exhibit a sudden drop at the transition. Examining the
density deviation profiles, we find that the high-$\ell$ modes are
edge modes, analogous to the surface modes in the 3D case. The
frequency drop of these edge modes at the hollowing transition
results from the redistribution of half of the radial nodes to the
emerging inner edge of the ring. Performing a similar analysis for
the edge modes in this disk geometry as we did for the surface
modes in the spherical geometry, we confirm that the degeneracy of
outer and inner edge modes in the hollow region is due to the
quadratic growth rate of the equilibrium density profile from the
center, a specific property of the bubble trap.

In summary, by considering a radially symmetric system in two
dimensions, we see that signatures of the hollowing transition
exhibited by the collective-mode spectrum are the same in both
three and two dimensions. This indicates that the signatures are
due to the physics of central hollowing, rather than details of
geometry or dimension. We remark that both disk-shaped and
ring-shaped BECs have been well studied
theoretically~\cite{Stringari2006,Perrin2012} and
experimentally~\cite{Ramanathan2011,Gupta2005}. The transition
regime between these 2D topologies is potentially achievable but
has not been studied, to our knowledge. Our findings not only
detail the collective-mode physics in the transition regime but
also reveal universal features in hollowing-out condensate
systems.

\section{Effects of Gravity}\label{sec:VII}

So far we have examined the equilibrium profiles and dynamical
behavior of spherically symmetric BECs as their spatial topology
is changed from filled to hollow. We now discuss gravitational
effects which cannot be neglected in experimental shell traps on
Earth. Crucially, unlike in a harmonically trapped system, in a
shell-shaped trapped system the gravitational force tends to cause
sag: mass accumulation at the lower vertical points in the system
and a depletion around the highest. This gravitational sag has
been experimentally shown to produce quasi-2D systems having no
closed shell-like surfaces under ordinary gravitational conditions
on Earth. We estimate the gravitational strength at which the
density of the top region of the condensate becomes completely
depleted for a thin-shell geometry within the Thomas-Fermi
approximation and arrive at a precise value for the critical
number of atoms to produce a closed thin-shell structure. Finally,
in a perturbative treatment, we analyze the effect of low gravity
on the collective mode structure of a very thin condensate shell.
We find that gravity couples modes with adjacent angular momentum
indices. This is consistent with the fact that a condensate shell
in a gravitational field is not fully spherically symmetric. Our
estimates show that for typical cold atomic experimental settings,
microgravity facilities are the most promising for studying the
rich collective mode structure of closed condensate shells.

\subsection{Behavior of equilibrium density and open shells} \label{sec:VIIA}

First, we identify the condition for treating gravity as a small
effect compared to the strength of interactions between the atoms
in the condensate. We work in the strong interaction and thin
shell limits of the radially shifted harmonic potential discussed
in Sec.~\ref{sec:IVB}. This geometry can be achieved by the bubble
trap. In order to arrive at the needed condition, we consider the
case in which the effect of gravity shifts the trap minimum by an
amount much smaller than the thickness of the BEC shell, as
determined by the strength of interatomic interaction. More
precisely, as noted in Sec.~\ref{sec:IVB}, the thickness of the
condensate shell can be obtained from the Thomas-Fermi density
profile after fixing the number of particles, $N$, as was found in
Eq.~(\ref{eq:shellthick}),
\begin{equation*}
\delta=\frac{1}{2}\left(\frac{3UN}{m\pi\omega^2_{\rm{sh}}S_l^2r_0^2}\right)^{1/3}.
\end{equation*}
The displacement of the trap minimum away from its center
due to gravity~\cite{Scheinder1999} is equal to
\begin{equation}\label{eq:mindisplacement}
r_{\mathrm{disp}}=\frac{g}{\omega_{\rm{sh}}^2S_l}.
\end{equation}

\begin{figure}[t]
\centering
  \includegraphics[width=8.6cm]{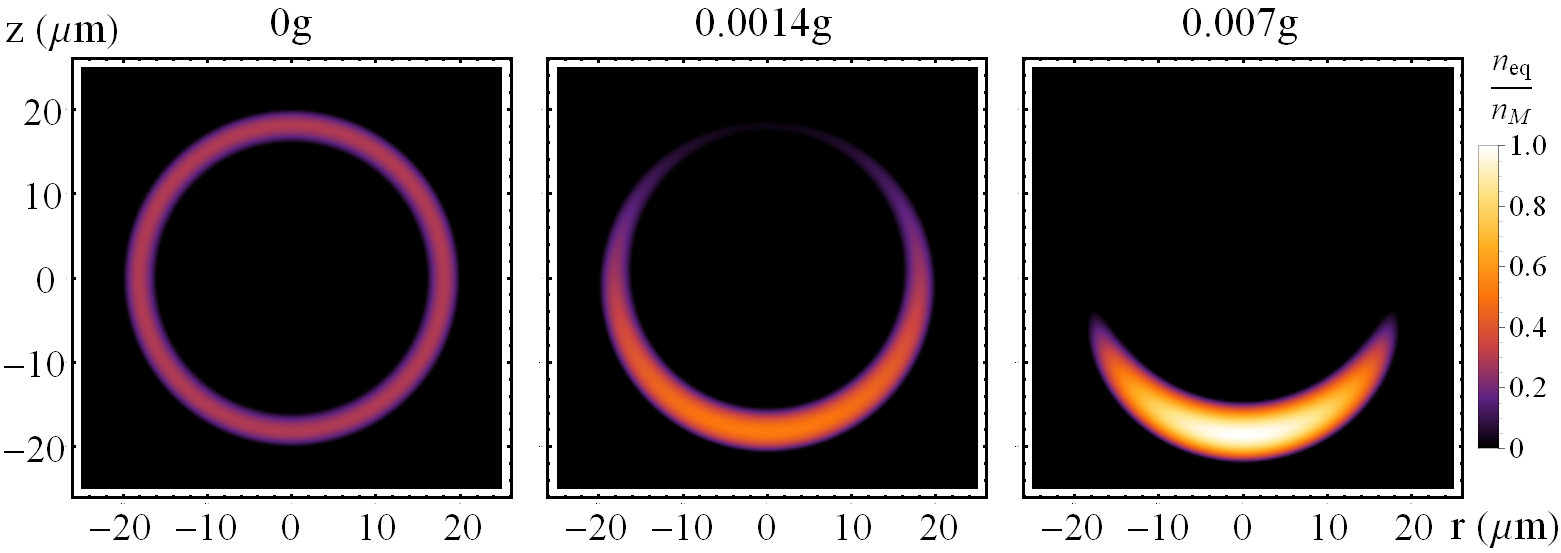}
\caption{(Color online) Thomas-Fermi density profiles for
condensates confined by the bubble trap without gravity (left) and
under the influence of gravitational fields $0.0014g$ (middle) and
$0.007g$ (right), where $g$ is the gravitational acceleration on
Earth. These profiles are generated for $10^5$ $^{87}$Rb atoms
forming a condensate shell with outer radius $20$ $\mu$m and
thickness $4$ $\mu$m in the absence of gravity. The colors in the
bar graph represent density normalized by $n_M=3.1 \times
10^{13}/\rm{cm}^{3}$. As the strength of the gravitational field
increases, we observe a density depletion at the top of the
condensate shell and a density maximum at its bottom.}
        \label{fig:13}
\end{figure}

Hence, the influence of gravity is small compared to the strength
of interactions in the condensate when
\begin{equation}\label{eq:weakcondition}
r_{\mathrm{disp}} \ll \delta.
\end{equation}
When this condition is satisfied, we expect the BEC shell to be
largely unchanged in shape despite the influence of gravity. In experimental efforts of
Refs.~\cite{Colombe2004,Merloti2013}, the quantity analogous to
Eq.~(\ref{eq:mindisplacement}) is reported to be comparable to the
entire size of the condensate cloud. In this case, the condition
for weak gravity is clearly violated and a radical, flattening, deformation of
the condensate shell is found by the authors.

As an intermediate stage between these two possibilities---weak
gravity that does not change the shape of the condensate shell and
very strong gravity that effectively collapses it into a quasi 2D
geometry---we consider a set of experimental parameters for which
gravity deforms the BEC shell to have most density at its bottom
and a heavily depleted top region. This change in condensate
density is shown for the bubble trap and multiple values of
gravitational field smaller than Earth's gravity in
Fig.~\ref{fig:13}.

To examine the deformation analytically, we work within the
Thomas-Fermi approximation and consider the thin-shell limit of
the potential described by Eq.~(\ref{eq:Vsh}) with an added
gravitational term. Specifically, adding $-gr\cos\theta$ to the
potential and completing the square, we obtain
\begin{eqnarray}\label{eq:compltsq}
&&  V(r,\theta)=\frac{m\omega^2_{\rm{sh}}S_l^2}{2}\left[(r-r_0)
+\frac{g\cos\theta}{\omega_{\mathrm{sh}}^2S_l}\right]^2 \nonumber\\
&&  +\left(mgS_lr_0\cos\theta-\frac{mg^2\cos^2\theta}{2\omega_{\mathrm{sh}}^2}\right),
\end{eqnarray}
where $r_0$ is the trap minimum in the absence of gravity. For
the bubble trap, the frequency of small single-particle oscillations is given by
$\omega_{\rm{sh}}=\sqrt{\Delta/\Omega}\omega_{0}$ and the trap minimum is $r_0=\sqrt{\Delta}$.

The potential in Eq.~(\ref{eq:compltsq}) is equivalent to a
radially shifted harmonic trap in the variable $r-r_0$, with the
trap minimum vertically displaced away from the center of the
condensate shell. This vertical displacement is equivalent to
adding a potential term
\begin{equation}\label{eq:Voffset}
V_{\rm{offset}}(\theta)=mgS_lr_0\cos\theta-\frac{mg^2\cos^2\theta}{2\omega_{\rm{sh}}^2}.
\end{equation}
From the Thomas-Fermi equilibrium density, Eq.~(\ref{eq:neq}), we
see that the density at the top of the shell vanishes when the
chemical potential is equal to $\mu_c=V_{\rm{offset}}(\theta=0)$.
This corresponds to the condensate shell ``opening up'' and can,
in extreme cases, lead to the shell approaching a quasi-2D
geometry as in Refs.~\cite{Colombe2004,Merloti2013}.

When the chemical potential attains the critical value $\mu_c$,
the inner and the outer radii of the shell are the solutions of
$\mu_c-V(r)=0.$ Previously, when the effects of gravity were not
take into account, we typically fixed the outer radius of the
condensate as $r=R$, with $R$ being some constant chosen in
accordance with experimental data. Once gravity is accounted for,
the condensate shell is not perfectly spherically symmetric, so
there is no longer a single, $\theta$-independent, outer radius
$R$. Similarly, in the presence of gravity, the thickness of the
shell varies with the polar angle as
\begin{equation}\label{eq:thicknesstilde}
\delta(\theta)=2r_0\sqrt{\tilde{g}(1-\cos\theta)(2-\tilde{g}-\tilde{g}\cos\theta)},
\end{equation}
where we have introduced the dimensionless parameter
$\tilde{g}=g/S_lr_0\omega_{\rm{sh}}^2$. We take the condensate
shell corresponding to the potential of Eq.~(\ref{eq:compltsq}) to
be very thin for $\tilde{g} \ll 1$ and refer to these values of
$\tilde{g}$ as the thin shell limit in analogy with our
discussion in Sec.~\ref{sec:IVB}.

We note that $\delta(0)=0$, as expected since we have chosen the
chemical potential to cause a density depletion at the top of the
condensate shell. Additionally, for $\theta=0$ we find
$r_{\rm{max}}=r_{\rm{min}}=r_0-\frac{g}{\omega_{\rm{sh}}^2S_l}$
and since these values should correspond to coordinates on top of
the BEC shell we require that
\begin{equation}\label{eq:SmithaConstraint}
r_0>\frac{g}{\omega_{\rm{sh}}^2S_l}.
\end{equation}
This is a rather natural constraint given that in the classical
problem of an oscillator in a gravitational field,
the frequency of oscillation is given by
$\omega=\sqrt{g/L}$ for $L$ the size of the object that is
oscillating. In other words, the condition in
Eq.~(\ref{eq:SmithaConstraint}) requires the trapping frequency to
be larger than the oscillation frequency naturally associated with
gravitational effects. These observations fully characterize the
geometry of the condensate shell deformed by gravity.

Furthermore, as is relevant in experimental settings, total
particle number conservation requires
\begin{eqnarray}\label{eq:totno}
N_c=2\pi\int_{r_{\rm{min}}}^{r_{\rm{max}}}\int_0^{\pi}n_{eq}(r,\theta)r^2
\sin \theta drd\theta,
\end{eqnarray}
which, evaluated in the thin shell limit, is equivalent to
\begin{equation}\label{eq:Nestimate}
  N_c=\frac{128m\pi g^{3/2}r_0^{7/2}S_l^{1/2}}{15U\omega_{\rm{sh}}}.
\end{equation}
Since this expression only includes the gravitational acceleration
and trap parameters (the minimum of the harmonic trapping
potential $r_0$ and the frequency of single-particle oscillations
around the trap minimum $\omega_{{\rm{sh}}}$), it can be used to
predict at what particle number the effects of gravity will cause
a density depletion at the top of the condensate shell.

Accordingly, given fixed trap parameters, a value of $N$ can be
chosen in such a way that the thin condensate shell keeps an
approximately uniform density profile regardless of gravitational
effects: Choosing a number of atoms larger than $N_c$  guarantees
a smaller deformation of the shape of the BEC shell under the
influence of gravity. In the limit of $g\to 0$ the number of atoms
needed in order to form a condensate shell robust to the influence
of gravity, $N_c$, vanishes and it also diminishes as
$\omega_{\rm{sh}}$, the trapping frequency, is increased. In other
words, an equivalent of the weak gravity regime can be reached by
using tighter confinement for the trapped atoms as well.

Furthermore, Eq.~(\ref{eq:Nestimate}) implies that
given a fixed number of atoms forming the condensate, a critical
value of the gravitational constant $g_c$ can be estimated as
\begin{equation}\label{eq:gc}
  g_c=\frac{(15U\omega_{sh}N)^{2/3}}{(128m\pi)^{2/3}r_0^{7/3}S_l^{1/3}}.
\end{equation}
In experimental setups where the measured value of $g$ is larger
than $g_c$, the condensate shell is expected to open up on top,
rather than maintaining its bubble-like shape, as depicted in
Fig.~\ref{fig:13}. If the effective value of the gravitational
constant can be lowered through experimental techniques, or use of
special microgravity facilities, we propose that realizing
$g_{{\rm{eff}}}$ smaller than $g_c$ will result in a robust
shell-like geometry of the condensate.

Under experimental conditions within which the value of
gravitational acceleration cannot be changed, the expression in
Eq.~(\ref{eq:Nestimate}) suggests a way in which shell-shaped
condensates can also be achieved and maintained without
significant gravitational sag. In cases where $N\gg N_c$ is within
the range of plausible experimental parameters, gravity can be
treated as a small effect on an otherwise stable condensate shell
of approximately uniform thickness. We proceed to examine the
collective motions of condensate shells under such conditions.

\subsection{Collective modes in presence of gravity} \label{sec:VIIB}

We turn to a discussion of condensate shell dynamics in the
presence of gravity. In particular, we consider including gravity
in the hydrodynamic formalism. Within this approach, we use the
gravitational GP equation and apply the Thomas-Fermi approximation
with gravity explicitly accounted for.

We start by considering an addition of the gravitational potential
term to Eq.~(\ref{eq:GPeqn}) so that
$V({\bf{r}})=V_{{\rm{trap}}}({{r}})+mgz$ and consequently
obtain
\begin{eqnarray}\label{eq: GPg1}
&& \frac{-\hbar^2}{2m}\nabla^2\psi({\bf{r}},t)+V_{{\rm{trap}}}({{r}})\psi({\bf{r}},t)-mgz\psi({\bf{r}},t)\nonumber\\
&& +U_0|\psi({\bf{r}},t)|^2\psi({\bf{r}},t)=i\hbar\partial_t\psi({\bf{r}},t),
\end{eqnarray}
a GP equation with an explicit gravitational term. Further,
the corresponding Thomas-Fermi density profile reads
\begin{equation}\label{eq:TFgrav21}
n_{\rm{eq}}({\bf{r}})=\frac{\mu-V_{{\rm{trap}}}({{r}})-mgz}{U}.
\end{equation}

The hydrodynamic equations appropriate for a BEC in a
nonzero gravitational field are then
\begin{eqnarray}\label{eq:hydroFin1}
&& \omega^2\delta n({\bf{r}})=\frac{1}{m S_l^2}\frac{\partial V_{{\rm{trap}}}(r)}{\partial r}\frac{\partial \delta n({\bf{r}})}{\partial
 r} + \frac{g}{S_l}\cos\theta\frac{\partial \delta n({\bf{r}})}{\partial
 r}\nonumber\\
 && -\frac{g}{r S_l}\sin\theta\frac{\partial \delta n({\bf{r}})}{\partial
  \theta} - \left[\frac{\mu-V_{{\rm{trap}}}(r)}{m S_l^2}-\frac{gr}{S_l}\cos\theta\right]\nabla^2\delta
  n({\bf{r}}).\nonumber\\
\end{eqnarray}
This expression, Eq.~(\ref{eq:hydroFin1}), is the nonzero gravity
equivalent of Eq.~(\ref{eq:delta_n_no_TF2}) with the Thomas-Fermi
equilibrium density profile, Eq.~(\ref{eq:TFgrav21}). As in
previous sections, we use dimensionless length units rescaled by
$S_l=\sqrt{\hbar/(2m\omega)}$.

We proceed to treat the gravitational terms in this equation
perturbatively: We assume that the trap parameters have been
chosen in accordance with Eq.~(\ref{eq:Nestimate}) so that the
shape of the shell is largely unchanged by gravitational effects.
Additionally, we note that choosing $N\geq N_c$ satisfies the
condition in Eq.~(\ref{eq:weakcondition}), thus justifying our
treatment of the condensate shell thickness as effectively
uniform.

As shown in Sec. \ref{sec:IVB}, for a very thin condensate shell
the collective modes are analytically described by
\begin{eqnarray}\label{eq:deltan}
\delta
n({\bf{r}})=\sqrt{\frac{\nu(\nu+1)}{2}}P_{\nu}\left(\frac{r-r_0}{\delta}\right)Y^l_{m_\ell}(\theta,\phi),
\end{eqnarray}
where $P_{\nu}(x)$ are the Legendre polynomials, with the
corresponding frequencies given by
Eq.~(\ref{eq:thin_shell_modes}),
\begin{equation*}
\omega_{\nu,\ell}^{\rm{sh}}=\omega_{\rm{sh}}\sqrt{\nu(\nu+1)/2}.
\end{equation*}
Corrections to these frequencies due to the effects of gravity are
then given by the eigenvalues of a matrix with entries equal to
\begin{eqnarray}\label{eq:matrixelt}
&& \langle \delta
n({\bf{r}})_{\nu,m_{\ell}}^{\ell}|V_g(r,\theta)|\delta
n({\bf{r}})_{\nu',m'_{\ell}}^{\ell'}\rangle = \frac{g}{S_l}\langle
\delta
n({\bf{r}})_{\nu,m_\ell}^{\ell}|\nonumber\\
&& -\frac{1}{r}\sin\theta\frac{\partial}{\partial \theta}
+\cos\theta\frac{\partial}{\partial r}+r\cos\theta\nabla^2|\delta
n({\bf{r}})_{\nu',m'_{\ell}}^{\ell'}\rangle.
\end{eqnarray}
In the thin-shell limit,  $c \gg 1$, we only consider the matrix
elements of Eq.~(\ref{eq:matrixelt}) to leading order in $c^{-1}$,
or equivalently, the thickness of the condensate shell, $\delta$.
Carrying out this calculation (see more details in Appendix
\ref{app:matrixelt}), we find
\begin{eqnarray}\label{eq:matrixeltfinp}
&& \langle \delta
n({\bf{r}})_{\nu,m_{\ell}}^{\ell}|V_g(r,\theta)|\delta
n({\bf{r}})_{\nu',m'_{\ell}}^{\ell'}\rangle\approx \frac{g\delta
r_0}{S_l}\frac{\nu(\nu+1)}{2\nu+1}  \nonumber\\
&& \times [f(\ell,\ell',m_{\ell},m_{\ell'})\delta_{\ell,\ell'+1}+
g(\ell,\ell',m_{\ell},m_{\ell'})\delta_{\ell,\ell'-1}],
\end{eqnarray}
where the numerical factors $f(\ell,\ell',m_{\ell},m_{\ell'})$ and
$g(\ell,\ell',m_{\ell},m_{\ell'})$ are given in Appendix
\ref{app:matrixelt}. Noting the Kronecker $\delta$ functions in
Eq.~(\ref{eq:matrixeltfinp}), we conclude that for a fixed $\nu$
and $\ell$ finding the eigenfrequencies and eigenmodes of the
system under the influence of gravity is reduced to diagonalizing
a matrix with nonzero entries only for
\begin{equation}\label{eq:modemix}
\{\nu',\ell',m'_{\ell} \}=\{\nu,\ell \pm
1,m_{\ell} \}.
\end{equation}
Consequently, the effect of weak gravity (gravity in the regime
where it can be treated perturbatively) on the collective modes of
the spherically symmetric thin condensate shell is to mix modes
with adjacent angular momentum indices. Therefore, if a collective
mode with a fixed number of radial nodes $\nu$ and a fixed number
of angular modes $\ell$ is induced under conditions of weak
gravity in a thin shell BEC, the number of angular nodes will
change while the radial density-deviation profile remains the
same. Since the overlap between the collective modes under
consideration, Eq.~(\ref{eq:matrixeltfinp}) is weighted by both
the thickness of the shell and the minimum of the harmonic
trapping potential, the mixing effect will be more or less
prominent depending on the size of the condensate shell. Further,
the same conclusion of gravitationally induced mode mixing can be
obtained by making an ansatz $\delta
n(r,\theta,\phi)=D(r)\sum_{\ell=0}^{\infty}\sum_{m_{\ell}=-\ell}^{\ell}C_{\ell
m_\ell}Y^{\ell}_{m_\ell}(\theta,\phi)$, for $C_{\ell m_\ell}$ some
appropriate set of constants, and seeking a complete solution to
the eigenproblem of Eq.~(\ref{eq:hydroFin1}).

Finally, we emphasize that Eq.~(\ref{eq:matrixeltfinp}) is
obtained not only in the weak gravity regime but also the thin
shell limit so that the mode mixing effect on a thicker shell
would be qualitatively different than on a thin one. To gain
insight into the behavior of a thicker shell described by a
smaller value of $c$ (larger $\delta$), we note that evaluating
the matrix element of Eq.~(\ref{eq:matrixeltfinp}) to quadratic
order (see Appendix \ref{app:matrixelt} for more details) in the
shell thickness yields terms such as
\begin{eqnarray}
 2\delta^2\left(\begin{array}{ccc}
 \nu & 1 & \nu' \\
  0 & 0 & 0
 \end{array}\right)^2,
\end{eqnarray}
where we use the Wigner-$j$ symbol. As this symbol is proportional
to a Clesbsh-Gordan coefficient, we can identify a selection rule
for this expression. More precisely, terms of this form vanish
unless $\nu'=\nu\pm 1$. Consequently, we posit that away from the
thin-shell limit (assuming that gravity can still be treated as
weak compared to the strength of interactions between the atoms
making up the condensate), gravitational pull leads not only to
the mixing of modes with adjacent angular node indices $\ell$ but
also those with adjacent radial node indices $\nu$. The practical
impact of the analysis of gravitational effects presented in this
section is discussed below.

\section{Applications And Experimental Feasibility}\label{sec:VIII}

Having presented an extensive study of spherically symmetric
hollow condensates and their behavior in various limits, we turn
to a discussion of experimental feasibility of achieving these
configurations. We start by providing a few estimates for
experimentally measurable quantities, such as condensate density
and collective mode frequency, then turn to the effects of
gravity, given its striking influence on shell-shaped BECs on
Earth.

As a realistic example, we consider a fully filled spherical
$^{87}$Rb condensate made up of $N=10^5$ atoms and created by the
potential of Eq.~(\ref{eq:harmonic}) with the bare frequency
$\omega=2\pi\times 500$ Hz and condensate size of $R=10 \
\mu\mathrm{m}$.  The maximum density of such a system is on the
order of $10^{16}$ cm$^{-3}$ while for a condensate shell of the
same size and equivalent confining frequency with
$c=r_0/\delta\approx 1000$, we find that the maximum condensate
density is on the order of $10^{18}$ cm$^{-3}$. Comparing their
respective breathing mode frequencies shown in Fig.~\ref{fig:3} we
find that the $(\nu,\ell)=(1,0)$ mode in the filled condensate
corresponds to $\omega^{\mathrm{sp}}_{1,0}\approx 2\pi\times 1.12
$ kHz while the same mode in the thin shell is characterized by
$\omega^{\mathrm{sh}}_{1,0}\approx 2\pi\times 0.50$ kHz with the
correction given by Eq.~(\ref{eq:thinshellpert}) on the order of
$10^{-6}\%$ and therefore negligible. In other words, recalling
our discussing in Sec.~\ref{sec:VA} and Fig.~\ref{fig:3}, we
predict that when the $(\nu,\ell)=(1,0)$ mode is induced in the
fully filled spherical condensate, if the system transitions to a
thin shell, its eigenfrequency decreases from $2\pi\times 1.12$
kHz to a value slightly smaller than $2\pi\omega\approx 2\pi\times
0.50$ kHz at the hollowing transition and then increases to reach
this value in the very-thin-shell limit. The adiabatic change in
the condensate shape from a filled sphere to a very thin, hollow
shell therefore results in approximately halving the lowest lying
collective mode frequency. Similarly, we calculate
$\omega^{\mathrm{sp}}_{2,0}\approx 2\pi\times 1.87$ kHz and
$\omega^{\mathrm{sh}}_{2,0}\approx 2\pi\times 0.87$ kHz, a $53\%$
decrease from the fully filled sphere limit to the thin hollow
shell geometry. More generally, for all low-lying, $\ell=0$ modes
that are experimentally accessible, we predict the decrease in the
collective mode frequency at the hollowing point, compared to the
oscillation frequency of the same mode in the fully filled
spherical BEC to be rather prominent, on the order of $50\%$ or
more. Somewhat higher spherically symmetric modes, such as
$\nu=3$, are good candidates for experimental detection of the
further decrease of collective mode frequency at the hollowing
point compared to the thin-shell limit as well. The collective
mode with $\nu=3$ shows a $20\%$ change between the hollowing
point and the thin-shell limit, which makes it suitable for full
observation of the nonmonotonicity of the collective mode
frequency spectrum of a hollowing BEC.

Additionally, we note that collective modes with low angular
momentum values, such as $\ell=1$ or $\ell=2$, exhibit frequency
dip features similar to those in frequencies of spherically
symmetric $\ell=0$ modes. In a realistic experimental system that
might not have perfect spherical symmetry, such low-$\ell$
collective modes would be the most likely candidate of study.

Comparing the modes with $\nu=1$ but different $\ell$, we find
that $\omega^{\mathrm{sp}}_{1,1}\approx 2\pi\times 1.41$ kHz while
$\omega^{\mathrm{sp}}_{1,10}=2\pi\times 2.96$ kHz and
$\omega^{\mathrm{sp}}_{1,20}=2\pi\times 4.03$ kHz. Identifying
large $\ell$ modes, such as $\ell=10$ and $\ell=20$, with
excitations of the condensate localized to a particular boundary
surface, we note that these surface modes typically have high
oscillation frequencies compared to modes with the same number of
radial nodes $\nu$ but fewer oscillations in the angular
direction. Given our calculations in Sec.~\ref{sec:IVB}, this
effect of increasing $\ell$ is not present in the thin-shell limit
since all of the collective modes in this configuration are
degenerate with respect to a change in the number of their angular
nodes.

For $\ell=20$, we further examine the drop in collective mode
frequency due to the redistribution of surface modes when an inner
boundary is available. Noting that collective modes with total
number of radial zeros equal to $2\nu$ and $2\nu+1$ become
degenerate immediately after the hollowing out transition takes
place, we find that $\omega^{\mathrm{sp}}_{1,20}=2\pi\times 4.03$
kHz and $\omega^{\mathrm{sp}}_{0,20}=2\pi\times 2.24$ kHz in the
sphere limit drop to $\omega^{\mathrm{sh}}_{1,20}=2\pi\times 1.50$
kHz. This drop constitutes a decrease on the order of $60\%$ and
$30\%$, respectively. Consequently, we identify the redistribution
of surface modes as a rather strong effect on collective mode
frequencies.

Finally, repeating our analysis for a thicker shell with $c\approx
20$, we find that the corrections to the collective mode
frequencies of the shell due to its nontrivial thickness become
approximately $3\%$ of the value obtained while assuming the
thin-shell condition $c\to\infty$ is satisfied. As this is not a
very large correction, we conclude that a number of our thin-shell
results can potentially be observed in experiments even when
creating a very thin condensate shell is not feasible.

We turn now to the effects of gravity in experimentally realistic
systems. In contrast to fully filled spherical condensates that
are simply shifted by gravity with no effect on their shape, the
geometry of a hollow condensate shell is, as we have shown in
Sec.~\ref{sec:VII}, rather sensitive to the presence of a
gravitational field. Accordingly, experimental parameters such as
the trapping frequency, interatomic interaction strength, and the
number of atoms in the condensate have to be chosen to satisfy the
criteria given by Eq.~(\ref{eq:Nestimate}); otherwise, the density
of the condensate shell is heavily depleted on its top while the
majority of atoms pool at its bottom.

In the presence of ordinary gravity, Eq.~(\ref{eq:Nestimate})
constrains the shell radius to be rather small or the trapping
frequencies to be rather high. Consequently, a realization of a
robust, fully shell-shaped BEC on Earth presents an experimental
challenge. We estimate that a thin condensate shell made up of
$^{87}$Rb atoms confined by a trap with a bare frequency of
$2\pi\times 500$ Hz and size on the order of $R \sim 10 \mu$m
would have to be made up of $N_c \sim 10^7$ atoms in order to
maintain approximately uniform density and robust shape under the
influence of gravity. As current experimental efforts
\cite{Lundblad,Hollow} show that even lower bare frequencies are
optimal for experiments using an rf-dressed bubble trap, we
conclude that the number of atoms needed in order for the
condensate shell to maintain its shape despite gravitational sag
is impractically high. In other words, experimental investigations
of the behavior and properties of $^{87}$Rb BEC shells in Earth's
gravity are not feasible and microgravity environments need to be
sought. For example, we estimate that for a $^{87}$Rb condensate
shell composed of 200,000 atoms, as is the case in
Ref.~\cite{Merloti2013}, of the same size ($R \sim 10 \mu$m) would
show significant lack of density on its top and largely consist of
a pool of atoms at the trap bottom unless (effective)
gravitational acceleration in its environment was smaller than
$g_c\sim 0.2$ m/$\rm{s}^2$.

While gravitational sag is often compensated for in experimental
studies of ultracold atomic systems on Earth, the shell-shaped
geometry makes standard methods such as magnetic levitation or the
use of the dipole force challenging. Since the bubble trap employs
a dressed-state potential, magnetic levitation is not feasible
because the atoms in the condensate are in superpositions of all
internal magnetic states. Further, though a dipole force due to a
gradient of optical intensity (for instance, achieved by using a
far-detuned Gaussian beam) could in principle be used to
counteract the effects of gravity, the need for high precision in
designing such a gradient makes this approach very difficult.
Consequently, we identify microgravity environments, rather than
terrestrial setups with gravitational compensation, as an optimal
choice for experimentally realizing hollow BEC shells.

In cases where a thin shell can be produced, most likely in
microgravity environments such as the ZARM drop tower~\cite{ZARM}
and the Matter-Wave Interferometry in Microgravity (MAIUS)
sounding rocket~\cite{MAIUS} in Germany, and NASA's Cold Atom
Laboratory aboard the International Space Station~\cite{CAL}, we
also estimate the corrections to the frequencies of the collective
modes due to their mixing under the influence of gravity. While
the magnitude of these corrections depends on the specific $\ell$
values of the modes being combined, we generally estimate them to
be rather small. In particular for the $\nu=1, \ell=1, m_\ell=0$
mode of a thin $^{87}$Rb condensate shell of the size on the order
of 1 $\mu$m and a bare frequency of $\omega_0=2\pi\times 500$ Hz
in microgravity $g\approx 10^{-5}$, we find the fractional
correction to the collective mode frequency $\omega$ to be
$\omega_{\rm{grav}}/\omega=1.00015$. This is a change in the
collective mode frequency of approximately 0.015 $\%$, which is
well below the detection limit in experiments. The magnitude of
gravitational mode mixing corrections increases in size with
increasing $\nu$, and similarly with increasing $\ell$. For
instance, for high-$\ell$ collective modes such as described by
$\nu=1$ and $\ell=20$ the effect of gravity leads to a 2.5 $\%$
change in collective mode frequency. As this is still a small
correction, we take the strength of gravity rather than the
magnitude of $\ell$ or $\nu$ to be the dominant factor for the
importance of mode mixing. In other words, the presence or absence
of mode-mixing effects in thin, spherically symmetric condensate
shells can be interpreted as an indicator of the strength of
gravity in experiments where a fully covered shell is realized.

More precisely, we posit that since the change in the frequencies
of the collective modes due to mode mixing is rather small in
microgravitational environments, the presence of large frequency
corrections due to mode mixing points toward a larger value of
gravitational acceleration. For instance, recalling that we
estimated that the condensate shell in Ref.~\cite{Merloti2013}
would retain its shape for $g_c\approx 0.2$ m/$\rm{s}^2$ we
calculate that the frequency of the $(\nu,\ell)=(1,0)$ breathing
mode in a thin shell of the same size ($R \sim 10 \mu$m) in a
gravitational field determined by $g_c$ would change approximately
$27\%$ due to mode mixing. This correction is almost 20,000 times
larger than it would be in microgravity. Additionally, a similar
mode-mixing behavior would be observed in condensate shells with
anisotropy along the $z$-direction, or clouds closer to a
``cigar-shape'' rather than a sphere. Therefore, the absence of
strong experimental signatures of mode mixing offers a
confirmation of both very small values of gravitational
acceleration in the environment of the condensate shell and also
its full spherical symmetry.

\section{Summary and Conclusion}\label{sec:IX}

Our comprehensive study of the collective excitations of hollow
BECs offers an understanding of quantum liquids having this unique
spatial topology and provides timely predictions for its
experimental investigation. A variety of cold atomic trapping
potentials are equipped to realize the hollowing transition as a
function of a tuning parameter, corresponding to the vanishing of
density within the BEC's interior, the rf-dressed bubble-trap
potential being a prime candidate.  We have studied density
profiles of such BECs by employing appropriate trapping
potentials, including a generalized radially shifted potential
capable of realizing multiple models of interior density
depletion. A realistic description of the central density
reduction is especially important for the analysis of the
hollowing-out change of the condensate's real space topology.

We have focused on the  collective modes of spherically symmetric
BECs and the nonmonotonic features in their frequency spectra that
serve as signatures of this topological transition. We have
employed a combination of exact analytic and numerical
hydrodynamics and real-time GP equation simulations of
sudden-quench experiments in a complementary manner. The latter
serves as an \emph{in situ} simulation of experimental probe of
BEC collective modes thus closely connecting our work to realistic
and feasibile experimental procedures. We have found that thin
shell BECs exhibit quantized collective modes that show a
significant frequency splitting due to radial degrees of freedom
but are degenerate with respect to angular degrees of freedom.
Upon transitioning between filled and hollow topology, the
breathing mode frequencies drop to a minimum, and the
corresponding density deviations concentrate around the interior
point of vanishing density. Additionally, as the hollowing-out
occurs, the redistribution of radial nodes in the
high-angular-momentum surface modes from the outer to the emerging
inner surface leads to a dramatic drop in their frequency
spectrum, constituting strong evidence of the hollowing-out
topological transition. We have additionally shown that these
nonmonotonic frequency features are robust across a range of
trapping potentials. Further, our analysis of the hollowing
transition of a 2D disk to a ring indicates that these breathing
and surface collective mode spectral signatures are universal for
hollowing-out topological transitions.

Finally, we have investigated the effects of gravity on a BEC
shell's ground-state stability and collective-mode frequencies. We
have determined critical experimental parameters for achieving
condensate shells that are not severely deformed by gravitational
sag. For terrestrial experimental conditions, gravity produces a complete
density depletion at the top of the shell potential
 and a pooling of atoms at the bottom of the shell,
which leads to a flattening of the 3D shell into a quasi-2D disk.
We argue that this sagging effect can be mitigated in microgravity
environments, and have made corresponding predictions for the
perturbative effects of gravity on the collective mode spectra.

While BECs have already been created in microgravity~\cite{ZARM},
a series of experiments employing tunable trapping potentials
capable of executing a hollowing transition in NASA's Cold Atom
Laboratory (CAL) aboard the International Space Station are
expected to be the first experimental realization and
investigation of fully closed BEC shells~\cite{CAL_NatureNews}.
Our predictions for the collective mode behavior of such shells
will be particularly advantageous for probing the hollowing-out
phenomena, as direct imaging cannot clearly identify the emergence
of an inner surface within a BEC's interior.

Having exhaustively characterized the behavior of spherically
symmetric hollow BECs, many questions concerning similar hollow
systems remain to be addressed in the future. For instance, in
NASA's CAL experiments, the effects of asymmetric trapping
potentials on the collective modes and the hollowing signatures
in, e.g., time-of-flight expansion will be of significant
interest. Another major direction concerning hollow shells would
entail rotation; resultant vortex formation and distribution in
these geometries would exhibit behavior dramatically different
from those observed in thin spheres and disk geometries. Our
results are also applicable to broader settings of condensate
shells in neutron stars, Bose-Fermi mixtures, and
Mott-insulator-superfluid coexisting systems; future work would
require making concrete connections between these settings and our
analyses.

\begin{acknowledgments}
We thank Nathan Lundblad and Ryan Wilson for illuminating
discussions. KS acknowledges support by ARO (W911NF-12-1-0334),
AFOSR (FA9550-13-1-0045), NSF (PHY-1505496), and Texas Advanced
Computing Center (TACC). CL acknowledges support by the National
Science Foundation under Award No.~DMR-1243574. KP, SV, and CL
acknowledge support by NASA (SUB JPL 1553869 and 1553885). CL and
SV thank the KITP for hospitality.
\end{acknowledgments}

\appendix

\section{Finite-difference method} \label{app:Finite_diference}
In this section, we present the details for the finite-difference
method we use to solve the differential equation for the
collective modes. The Sturm--Liouville equation has a general form
of
\begin{eqnarray}
[p(x)y']' + q(x)y =  - \lambda w(x)y \label{eq:Sturm--Liouville},
\end{eqnarray}
where the eigenvalues $\lambda$ and corresponding eigenfunctions
$y(x)$ are to be determined. With a small interval $\varepsilon$,
we can approximate the differential term as
\begin{eqnarray}
{\left. {py'} \right|_x} = p(x)\frac{{y(x + 0.5\varepsilon ) - y(x
- 0.5\varepsilon )}}{\varepsilon },\label{eq:diff_term_1}
\end{eqnarray}
and hence
\begin{eqnarray}
&& {\left. {(py')'} \right|_x} = \frac{{{{\left. {py'} \right|}_{x
+ 0.5\varepsilon }} - {{\left. {py'} \right|}_{x - 0.5\varepsilon
}}}}{\varepsilon } \nonumber\\
&=& \frac{{p(x + 0.5\varepsilon )\frac{{y(x + \varepsilon ) -
y(x)}}{\varepsilon } - p(x - 0.5\varepsilon )\frac{{y(x) - y(x -
\varepsilon )}}{\varepsilon }}}{\varepsilon }\nonumber\\
 &=& \frac{p(x + 0.5\varepsilon )y(x + \varepsilon )}{\varepsilon ^2} - \frac{[ {p(x
+ 0.5\varepsilon ) + p(x - 0.5\varepsilon )}]y(x)}{\varepsilon ^2}
\nonumber\\
 && + \frac{{ p(x -
0.5\varepsilon )y(x - \varepsilon )}}{{{\varepsilon ^2}}}.
\label{eq:diff_term_2}
\end{eqnarray}
If the domain is sectioned into many lattice sites with lattice
spacing $\varepsilon$, a function $f$ can be represented by a
vector $\mathbf{f}= \{ f_i \}$ such as $f_i = f(x)$ and ${f_{i \pm
1}} = f(x \pm \varepsilon )$. We then turn
Eq.~(\ref{eq:diff_term_2}) to
\begin{eqnarray}
{(py')'_i} = \frac{{p_{i + 0.5}}{y_{i + 1}} - ({p_{i + 0.5}} +
{p_{i - 0.5}}){y_i} + {p_{i - 0.5}}{y_{i - 1}}}
{\varepsilon^2}.\nonumber\\ \label{eq:diff_term_3}
\end{eqnarray}
Using the rule of Eq.~(\ref{eq:diff_term_3}) and treating $q(x)$
and $w(x)$ as diagonal matrices, we can turn
Eq.~(\ref{eq:Sturm--Liouville}) into a generalized eigen problem
for a finite-size matrix. The accuracy can be increased by
decreasing $\varepsilon$. By comparing
Eq.~(\ref{eq:Sturm--Liouville}) with Eq.~(\ref{eq:eigenproblem})
multiplied by $r^2$, we obtain
\begin{eqnarray}
p &=& {r^2}\left[ {V(R) - V(r)} \right],\\
q &=& -l(l + 1)\left[ {V(R) - V(r)} \right],\\
w &=& {r^2},\\
\lambda &=& m \omega^2.
\end{eqnarray}
Therefore, one can calculate the collective modes by solving the
generalized eigen problem for a finite matrix.

\section{Evaluation of matrix elements in the perturbative approach to gravitational hydrodynamic equations}\label{app:matrixelt}

In this section, we present a detailed calculation of the matrix
elements of Eq.~(\ref{eq:matrixelt}). More precisely, to leading
order in the thickness of the condensate shell, we set out to
calculate
\begin{eqnarray}\label{eq:matrixeltCa}
&& \langle \delta n(r,\theta,\phi)_{\nu,m}^{l}|V_g(r,\theta)|\delta n(r,\theta,\phi)_{\nu',m'}^{l'}\rangle
\nonumber\\
&& \approx \frac{g}{S_l}\langle \delta n({\bf{r}})_{\nu,m}^{l}|-\frac{1}{r}\sin\theta\frac{\partial}{\partial \theta}
 -\frac{l'(l'+1)}{r}\cos\theta|\delta n({\bf{r}})_{\nu',m'}^{l'}\rangle.\nonumber\\
 \end{eqnarray}
To that end, we recall
 \begin{equation}\label{eq:deltanall}
 \delta n(r,\theta,\phi)_{\nu,m}^{l}=\sqrt{\frac{\nu(\nu+1)}{2}}P_{\nu}\left(\frac{r-r_0}{\delta}\right)Y^{l}_{m_\ell}(\theta,\phi)
 \end{equation}
 and, further, express the spherical harmonics as
 \begin{equation}\label{eq:Ylm}
 Y^{l}_m(\theta,\phi)=(-1)^m\sqrt{\frac{2l+1}{4\pi}\frac{(l-m)!}{(l+m)!}}P^{m}_l(\cos\theta)e^{im\phi}
 \end{equation}
 so that the matrix elements we are interested in require evaluating an integral
 over the polar and azimuthal angles and the radial coordinate.
 In order to evaluate the first term in Eq.~(\ref{eq:matrixeltCa}) we start by evaluating the integral over $\phi$
 as \begin{eqnarray}\label{eq:phiint}
 \int_0^{2\pi}e^{i(m-m')\phi}d\phi= 2\pi\delta_{m,m'}.
 \end{eqnarray}
  We proceed to utilize the differential identity
 \begin{eqnarray}\label{eq:thetadiff}
  &&\sin\theta\frac{\partial}{\partial\theta}P^{m}_l(\cos\theta)=\frac{1}{2l+1}[l(l-m+1)P^{m}_{l+1}(\cos\theta)\nonumber\\
  &&-(l+1)(l+m)P^{m}_{l-1}(\cos\theta)]
  \end{eqnarray}
  and the orthogonality relation
\begin{equation}\label{eq:ortho1}
  \int_{-1}^{1}P^m_k(\cos\theta)P^m_l(\cos\theta)\sin\theta d\theta=\frac{2(l+m)!}{(2l+1)(l-m)!}\delta_{k,l}
\end{equation}
for the integral over $\theta$. Furthermore, to evaluate the radial integral we carry
 out a change of variable $x=\frac{r-r_0}{\delta}$ and write
 \begin{eqnarray}\label{eq:matrixeltC1x}
  && \int P_{\nu}\left(\frac{r-r_0}{\delta}\right)P_{\nu'}\left(\frac{r-r_0}{\delta}\right)rdr\nonumber\\
  &&=\delta\int P_{\nu}(x)P_{\nu'}(x)(\delta x+r_0)dx
 \end{eqnarray}
 where the factor $\delta x+r_0$ can be expressed as a sum over the
Legendre polynomials $P_0(x)$ and $P_1(x)$. More precisely,
\begin{equation}\label{eq:legendreexp}
  \delta x+r_0=\delta P_1(x)+r_0P_0(x)
\end{equation}
so that the integral in Eq.~(\ref{eq:matrixeltC1x}) can be
evaluated by recalling that an integral of three Legendre
polynomials is proportional to the square of a Wigner-$j$ symbol
\begin{equation}
  \int_{-1}^1P_k(x)P_l(x)P_m(x)dx=2\left(\begin{array}{ccc}
 k & l & m \\
  0 & 0 & 0
 \end{array}\right)^2.
\end{equation}
The Wigner-$j$ symbol itself is related to a Clesbsh-Gordan
coefficient
\begin{equation}\label{eq:CGco}
  C(j_1,j_2,j_3|m_1,m_2,-m_3)=\langle j_1m_1,j_2m_2|j_3,-m_3\rangle
 \end{equation}
 by the definition
\begin{equation}\label{eq: wignerj}
  \left(\begin{array}{ccc}
 j_1 & j_2 & j_3 \\
  m_1 & m_2 & m_3
 \end{array}\right)=\frac{(-1)^{j_1-j_2-m_3}}{(2j_3+1)^{1/2}}C(j_1,j_2,j_3|m_1,m_2,-m_3).
\end{equation}
We can therefore evaluate the integral in Eq.~(\ref{eq:matrixeltC1x}) as
\begin{eqnarray}\label{eq:legendre3int}
 && \int P_{\nu}(x)(\delta x+r_0)
 P_{\nu'}(x)dx=\nonumber\\
&&\frac{2 r_0}{2\nu+1}\delta_{\nu,\nu'}+2\delta\left(\begin{array}{ccc}
 \nu & 1 & \nu' \\
  0 & 0 & 0
 \end{array}\right)^2
 \end{eqnarray}
Combining Eqs.~(\ref{eq:thetadiff}), (\ref{eq:matrixeltC1x}),
(\ref{eq:legendreexp}), and (\ref{eq:legendre3int}), we then
obtain, to leading order in $c^{-1}$,
\begin{eqnarray}\label{eq:matrixeltC1full}
   && \langle \delta n({\bf{r}})_{\nu,m}^{l}|\frac{g}{rS_l}\sin\theta\frac{\partial}{\partial \theta}|\delta n({\bf{r}})_{\nu',m'}^{l'}\rangle\approx\nonumber\\
  && \frac{g\delta
  r_0}{S_l}\frac{\nu(\nu+1)}{2\nu+1}\sqrt{\frac{(l'-m)!(l+m)!}{(2l+1)(2l'+1)(l'+m)!(l-m)!}}\nonumber\\
  &&\times \left[(l'+1)(l'+m)\delta_{l,l-1}-l'(l-m+1)\delta_{l,l'+1}\right].
\end{eqnarray}

We move on to calculate the second term in Eq.~(\ref{eq:matrixeltCa}). First, we note that the integral
over the azimuthal angle is equal to $2\pi\delta_{m,m'}$ as above. We then use
the recursive identity
\begin{equation}\label{eq:xP}
  xP^{l}_{m}(x)=\frac{l-m+1}{2l+1}P^{m}_{l+1}(x)+\frac{l+m}{2l+1}P^{m}_{l-1}(x)
\end{equation}
in order to evaluate the integral over the polar angle
\begin{equation}\label{eq:radialint2}
  \int P^{m}_{l}(\cos\theta)P^{l'}_{m}(\cos\theta)\cos\theta\sin\theta d\theta
\end{equation}
by again using the orthogonality relation of Eq.~(\ref{eq:ortho1}). We conclude that the total angular contribution to this term reads
\begin{eqnarray}\label{eq:matrixelt2}
  &&\int Y^{l'*}_{m}(\theta,\phi)Y^{l}_{m}(\theta,\phi)\cos\theta\sin\theta d\theta
  d\phi\nonumber\\
 &&=-\sqrt{\frac{(l+m)!(l'-m)!}{(2l+1)(2l'+1)(l-m)!(l'+m)!}}  \nonumber\\
 &&  \times \left [(l'-m+1)\delta_{l,l'+1} +(l'+m)\delta_{l,l'-1} \right ],
\end{eqnarray}
while the radial contribution is given by Eq.~(\ref{eq:legendre3int}). The total
contribution of this term is then equal to
\begin{eqnarray}\label{eq:2ndterm}
   && \langle \delta n({\bf{r}})_{\nu,m}^{l}|\frac{gl'(l'+1)}{rS_l}\cos\theta\frac{\partial}{\partial \theta}|\delta n({\bf{r}})_{\nu',m'}^{l'}\rangle\approx\nonumber\\
   && \frac{-g\delta r_0}{S_l}\frac{\nu(\nu+1)}{2\nu+1}\sqrt{\frac{(l'-m)!(l+m)!}{(2l+1)(2l'+1)(l-m)!(l'+m)!}}\nonumber\\
   && \times
   [l'(l'+1)(l'-m+1)\delta_{l,l'+1}+l'(l'+1)(l'+1)\delta_{l,l'-1}].
   \nonumber\\
\end{eqnarray}

We conclude that to leading order in $c^{-1}$, or equivalently the shell thickness $\delta$, the matrix element
of Eq.~(\ref{eq:matrixeltCa}) reads
 \begin{eqnarray}\label{eq:matrixeltfinapp}
    && \langle \delta n({\bf{r}})_{\nu,m}^{l}|V_g(r,\theta)|\delta
    n({\bf{r}})_{\nu',m'}^{l'}\rangle\approx \nonumber\\
  && \frac{g\delta r_0}{S_l}\frac{\nu(\nu+1)}{2\nu+1}\sqrt{\frac{(l'-m)!(l+m)!}{(2l+1)(2l'+1)(l-m)!(l'+m)!}}\times\nonumber\\
  &&\left[l'(l'-m+1)(l'+2)\delta_{l,l'+1}+(l'+m)(l'^2-1)\delta_{l,l'-1}\right].\nonumber\\
\end{eqnarray}

\end{document}